\documentclass[preprints,article,accept,moreauthors,pdftex]{Definitions/mdpi} 
\firstpage{1} 
\pubvolume{8}
\issuenum{5}
\articlenumber{290}
\pubyear{2022}
\copyrightyear{2022}
\externaleditor{Academic Editor: {Mauro D’Onofrio }}
\datereceived{15 April 2022}
\dateaccepted{17 May 2022}
\datepublished{22 May 2022}
\hreflink{https://doi.org/10.3390/{\linebreak}universe8050290} 
\pdfoutput=1

\Title{Gasflows in Barred Galaxies with Big Orbital Loops---A Comparative Study of Two Hydrocodes}

\TitleCitation{Gasflows in Barred Galaxies with Big Orbital Loops---A Comparative Study of Two Hydrocodes}

\Author{Stavros Pastras $^{{1,2,}\dagger}$, Panos A. Patsis $^{1,}$*$^{,\dagger}$ and E. Athanassoula $^{3}$}

\AuthorNames{Stavros Pastras, Panos Patsis and E. Athanassoula}

\AuthorCitation{Pastras, S.; Patsis, P.A.; Athanassoula, E.}

\address{$^{1}$ \quad Research Center for Astronomy and Applied Mathematics, Academy of Athens, RCAAM, Soranou Efessiou 4, 11527 Athens, Greece\\
$^{2}$ \quad Department of Physics, Section of Astrophysics, Astronomy and Mechanics, University of Athens, Zografos, 15784 Athens, Greece; spastras@phys.uoa.gr\\
$^{3}$ \quad Aix-Marseille Université, CNRS, CNES, LAM, Marseille, France; lia@lam.fr}

\corres{Correspondence: patsis@academyofathens.gr}

\firstnote{These authors contributed equally to this work.}

\abstract{We study the flow of gas in a barred-galaxy model, in which a 
considerable part of the underlying stable periodic orbits have loops where, 
close to the ends of the bar, several orbital families coexist and chaos 
dominates. Such conditions are typically encountered in a zone between the 4:1 
resonance and corotation. The purpose of our study is to understand the gaseous 
flow in the aforementioned environment and trace the morphology of the shocks 
that form. We use two conceptually different hydrodynamic schemes for our 
calculations, namely, the mesh-free Lagrangian SPH method and the adaptive mesh 
refinement code RAMSES. This allows us to compare responses by means of the 
two algorithms. We find that the big loops of the orbits, mainly belonging to 
the x1 stable periodic orbits, do not help the shock loci to approach 
corotation. They deviate away from the regions occupied by the loops, bypass 
them and form extensions at an angle with the straight-line shocks. Roughly at 
the distance from the center at which we start to observe the big loops, we find 
characteristic ``tails'' of dense gas streaming towards the straight-line 
shocks. The two codes give complementary information for understanding the 
hydrodynamics of the models.}

\keyword{galaxies: kinematics and dynamics;  barred; structure}
 
\begin{document}
%

\section{Introduction}
\label{introduction}

The general flow of the gas in a standard orbital environment, in which the 
orbits of the x1 family of periodic orbits (POs) have an elliptical shape with 
cusps or small loops, is described in \cite{a92b}. It has been shown that the 
strength of the bar component is, in turn, related with the shape of the 
dust lanes observed in the bars of the galaxies. This shape is associated with 
the morphology of x1 orbits, which constitute the backbone of the bar 
structure, as shown in the first part of this study, in \cite{a92a}.
However, the precise morphology of the underlying orbits is model-dependent and 
it may differ in its details. The loops of the x1 orbits can be large and so can 
the loops of the orbits of the families of the n:1 POs with n $\geqq 4$. In orbital 
models, POs belonging to x1 and POs belonging to the ``n:1'' families may be 
found in the same region of the disk. It is not clear how the flow of the gas in 
such regions can be adjusted.

Among the models in which we encounter big loops at the apocentra of the x1 and 
n:1 POs, are potentials estimated from near-infrared images of galaxies 
\cite{qfg94,boony,kpg10}, the general form of which can be written in polar 
coordinates $(r,\theta)$ as:
\begin{equation}\label{eq:potrthfour}
\Phi(r,\theta)=\Phi_0(r)+\sum_{k}\left[\Phi_{kc}(r)\cos(k\theta)
+\Phi_{ks}(r)\sin(k\theta)\right],
\end{equation}
where the components $\Phi_0(r)$, $\Phi_{kc}(r)$, and $\Phi_{ks}(r)$ 
are polynomials of the form $\sum_n \!a_n 
r^n$, with $k,n \in \mathbb{N}$
\cite{paq97,pkgb09,tp13} or cubic splines \cite{pkg10}. The polynomials, or the 
splines, are used to fit 
the observational data. All these models are also characterized by large chaotic 
regions, usually for Jacobi constants $E_J > E_J$ (4:1), although, in some cases, 
something such as this happens even for smaller energies \cite{pkg10}.

The main parameter that determines the evolution of the morphologies of the 
families of POs is the amplitude of the bar potential. Gaseous responses in 
cases where the bar-supporting orbits have loops with sizes of the order of 
a kiloparsec, are expected to be different than in cases where the orbits have 
smaller loops or even the orbits remain elliptical-like or even cuspy. However, 
there has been no study, as of yet, that investigated the details. This is attempted in the 
present paper.

A parallel goal of this study is to check how method-dependent are the predicted 
gaseous flows. Thus, we will also compare the responses of different hydrocodes 
by imposing the same potential, focusing always on the critical region between 
the 4:1 resonance and corotation. We will compare, between them, the 
gaseous responses of our models  using different methods, as well as with the 
results of the responses found by other authors. Nevertheless, the main goal of 
the paper is to associate the morphology of the shocks with the underlying 
orbital structure.
In Section~\ref{sec:potential} we describe the model that we  used in 
this study. In Section~\ref{sec:orbits}, we present the stellar response to 
our potential and identify the orbits that led to this. The gas 
responses by means of two different hydrocodes are given in 
Section~\ref{seq:gresponses}. In Section~\ref{discuss}, we discuss both the 
dependence of the results on the code used, as well as the relation between 
gaseous responses and orbital background; and, finally, we enumerate our 
conclusions in Section~\ref{conclusions}.


\section{The Model}
\label{sec:potential}

\begin{figure}[H]
\includegraphics[width=11 cm]{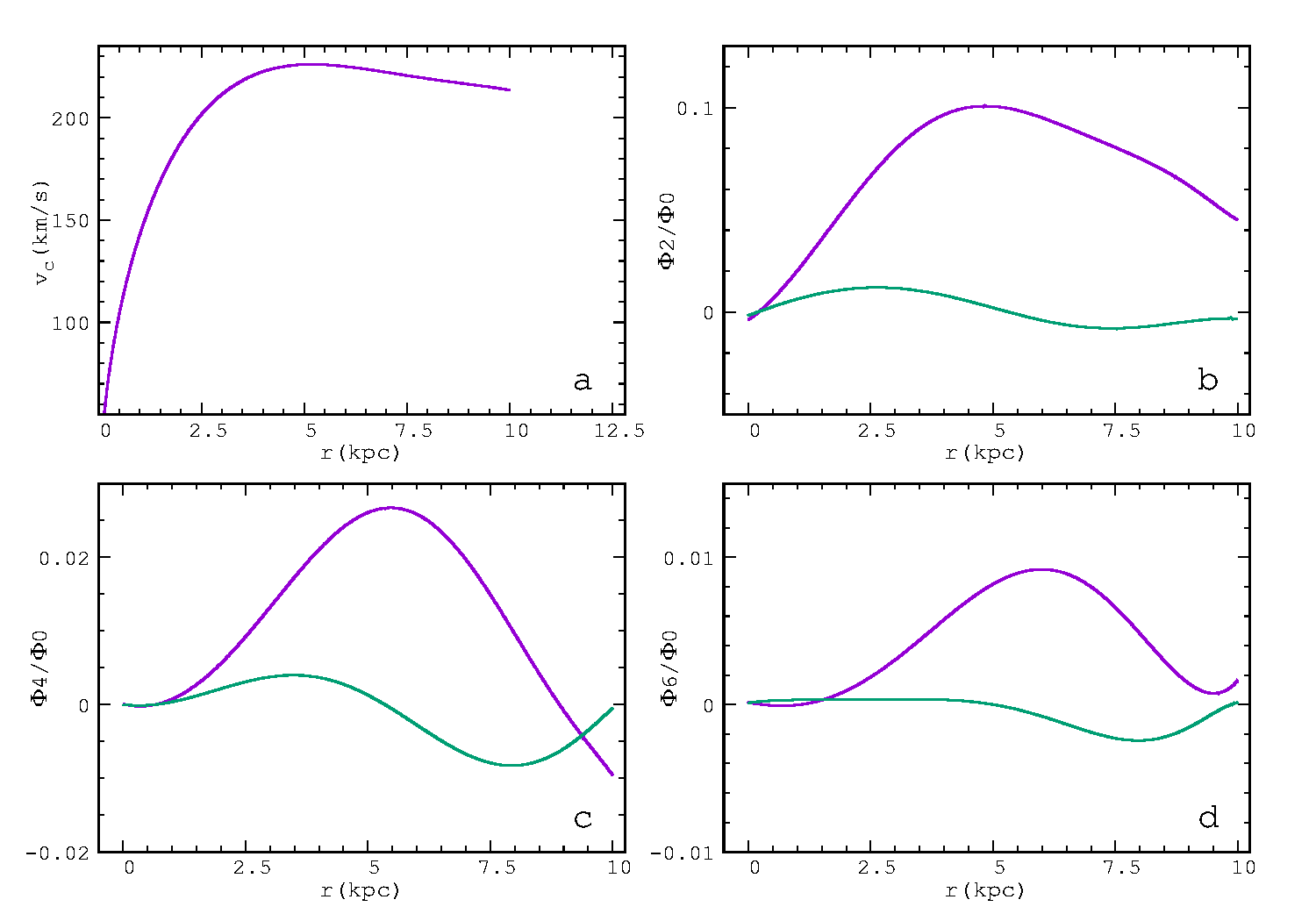}
\caption{The circular speed $v_c=\sqrt{r \frac{d\Phi_0}{dr}}$ (\textbf{a}) and the k = 2 
(\textbf{b}), k = 4 (\textbf{c}) and k = 6 (\textbf{d}) components of the potential, normalized by $\Phi_0$, 
as a function of the radius $r$. From (\textbf{b}--\textbf{d}), magenta curves refer to the 
cosine and green to the sine terms.
\label{rcpotcomp}}
\end{figure}

For our study, we chose a potential of the form of 
Equation~(\ref{eq:potrthfour}), with $k=2,4,6$, based on the estimation of the 
gravitational field of the barred galaxy NGC~7479 \cite{qetal95}. The adopted 
table with the $a_n$ coefficients, used in the fitting of the data with the 
$\sum_n\!a_n r^n$ polynomials, with $n=0,...8$, can be found in \cite{laine}. In 
Figure~\ref{rcpotcomp}a, we present the rotation curve, based on the $\Phi_0$ term, 
which is in agreement with the one given in Figure 8 in \cite{qetal95}. In 
panels b,c and d of Figure~\ref{rcpotcomp}, we give the variation in the $k=2,4,6$ 
components of the potential, respectively, normalized by $\Phi_0$. The cosine 
components are plotted with magenta, while the sine ones are the green color.

As we will see in Section~\ref{sec:orbits} below, this potential, 
rotating with a bar pattern speed $\Omega_b$, provides the orbital background 
against which we want to study the flow of gas. In addition, there are published
studies of gaseous responses, with which we can compare our results~\cite{laine,lsh98}.
In the present study, we did not try to model NGC~7479, or 
any particular galaxy. We, rather, investigated the flow of gas by two codes in 
the presence of a specific orbital background.

\section{Orbital Analysis}
\label{sec:orbits}

We derived the equations of motion from the autonomous Hamiltonian
\begin{equation}
\label{eq:hamilton}
H \equiv \frac{1}{2}\left(\dot{x}^{2} + \dot{y}^{2}\right) + \Phi(x,y) -
\frac{1}{2}\Omega_{b}^{2}(x^{2} + y^{2})=E_{J},
\end{equation}
where $(x,y)$ are coordinates in a cartesian frame of reference, rotating 
with angular velocity (pattern speed) $\Omega_{b}$. Furthermore, $\Phi(x,y)$ is the potential 
given in Equation~(\ref{eq:potrthfour}) and $E_{J}$ is 
the numerical value of the Jacobi constant. In the text, we will also refer to 
$E_{J}$  as the ``energy''. Dots denote time derivatives. In all 
our calculations, we used the following scaling of units $G=1$, $[L]=1$~kpc and 
$[v]=1$~km~s$^{-1}$. For the pattern speed, we adopted the value $\Omega_{b} 
=27$~km~s$^{-1}$~kpc$^{-1}$, proposed in \cite{laine, lsh98}, as the one matching the 
pattern speed of NGC~7479. In our calculations, the orientation of the bar was 
almost along the $y$-axis.

\subsection{Stellar Response Models}
\label{sresponses}

In order to understand the correspondence between the gas responses and the 
underlying stellar orbits in the model, we first ran a stellar response model 
and identified the orbits that shape its morphology.

We followed the same procedure for calculating the stellar response as the one we 
will follow in Section~\ref{seq:gresponses} for the gas models. Namely, we 
populated a $10$~kpc disk homogeneously with particles in a circular motion in the 
axisymmetric part of the potential, $\Phi_{0}(r)$ in \linebreak
Equation~(\ref{eq:potrthfour}), and we gradually increased the 
non-axisymmetric terms within a time corresponding to three 
pattern rotations $(3T)$, until we reached the full $\Phi(R,\theta)$ potential. 
Then, we continued integrating the particles for 10 pattern rotations more in the 
full potential. The positions and velocities of the particles at the end of the 
time-dependent part of the simulation were used as initial conditions for 
continuing their integration in the phase, during which our stellar response 
model behaved as an autonomous Hamiltonian system. We used a Runge--Kutta 4th-order scheme for the integration.

\begin{figure}[H]
\includegraphics[width=11.0 cm]{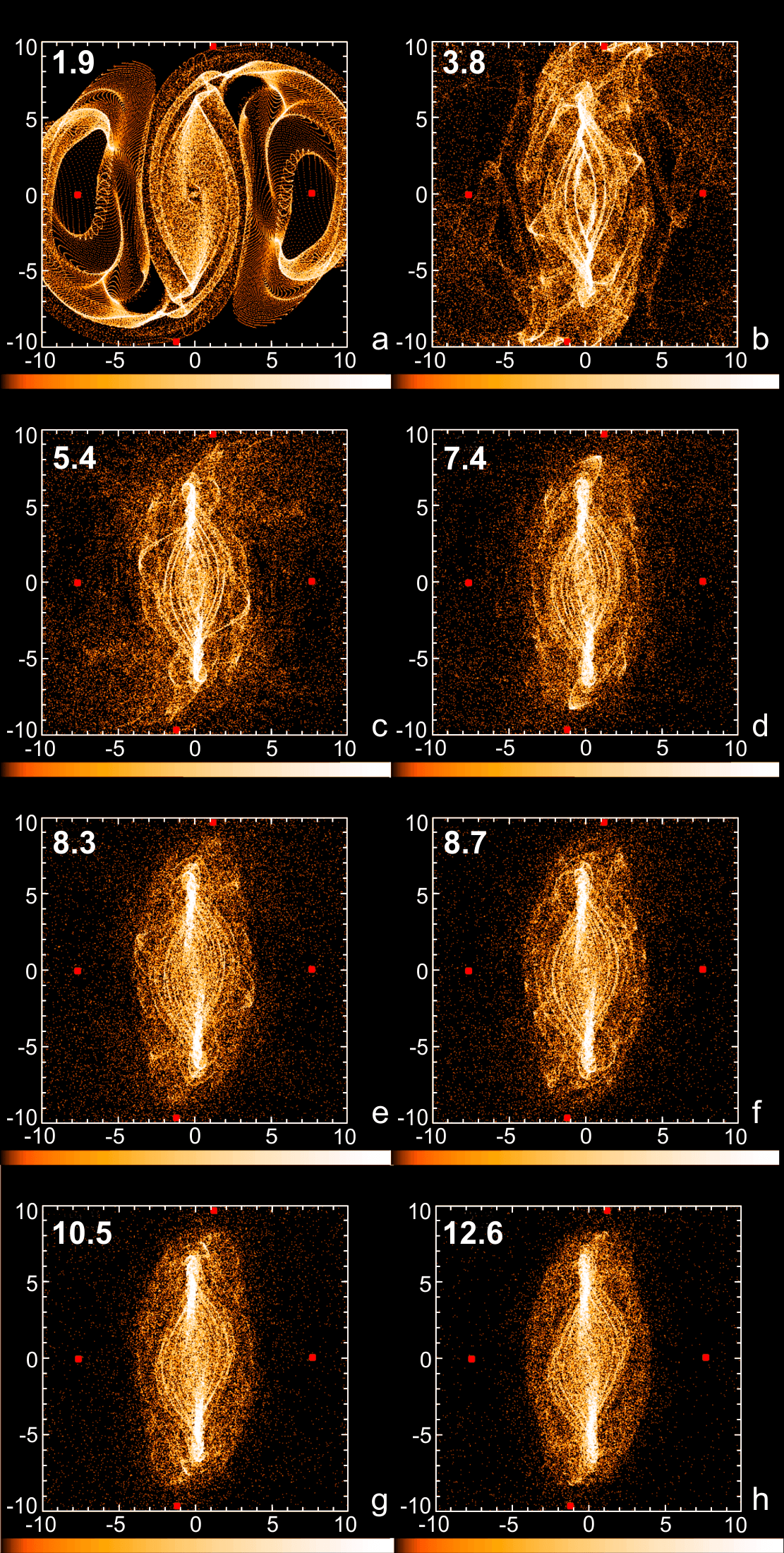}
\caption{(\textbf{a}--\textbf{h}): Successive snapshots of the stellar response model. 
Lighter regions correspond to denser regions, according to the logarithmic color bar at the 
bottom of the panels. The number of pattern rotations is given at the upper-left 
corner of the frames. The red dots indicate the location of L$_1$ (\textbf{top}),
L$_2$ (\textbf{bottom}), L$_4$ 
(\textbf{right}) and L$_5$ (\textbf{left}) of the bar. The bar rotates counterclockwise.
\label{stellarespo}}
\end{figure}

In Figure~\ref{stellarespo}, we give eight snapshots of the stellar response. Red 
dots indicate the Lagrangian points of the system. Those that are close to the 
ends of the bar are L$_1$, the upper, and L$_2$, the lower one, at 
$(x,y)=(1.21,9.65)$ and $(-1.21,-9.65)$, respectively, while those at the sides 
of the bar are L$_4$, to the right, and L$_5$, to the left side, at 
$(x,y)=(7.657,0.064)$ and $(-7.657,-0.064)$, respectively. The snapshots are 
converted to images by taking into account the number density of the particles 
in each pixel. The corresponding pixel intensity increases from left to right, 
according to the (logarithmic) color bars at the bottom of the panels, which 
span from min = 0 to max = 25.

The successive morphologies are given in the number of pattern rotations indicated 
in the upper-left corners of the panels. The transient, bar-growing, phase lasts up to $t=0.69=3T$. 
In all the snapshots, there are discernible structures reminiscent of x1 POs. In 
Figure~\ref{stellarespo}b--h, we observe a central, oval region reaching 
$y\approx 4.5$\,kpc. A dense straight-line feature, almost along the major axis 
of the bar, emerges through this oval region, reaching slightly larger distances 
from the center. Just outside the oval region, we can also observe loops 
(especially in Figure~\ref{stellarespo}c,e).

Another characteristic of the 
stellar response model is that the region with the features identified with the 
x1 POs is embedded within a more extended envelope. This component reaches a 
$x_{max}\approx 4$~kpc and a $y_{max}\approx 8$~kpc. Due to its presence, we can 
describe the resulting morphology as composed of two ``barred'' structures. An 
inner ``x1-bar'' and an outer envelope of the bar. Inside the latter, we can 
discern some ``corners'' of orbits, not always at the same place (e.g., in panels 
Figure~\ref{stellarespo}b,c), which eventually seem to be smoothed out in 
panels Figure~\ref{stellarespo}e,f. This is a direct indication of sticky 
chaotic orbits \cite{ch08} participating in its formation. Finally, we observe 
that the areas around L$_4$ and L$_5$ gradually become depleted of particles.

\subsection{Orbits}
\label{orbits}

\subsubsection{Periodic Orbits (POs)}
\label{porbits}

POs are periodic solutions of the system of differential equations derived from 
\linebreak Equation~(\ref{eq:hamilton}), i.e., from the equations of motion. POs do not exist in 
galaxies. Nevertheless, knowledge of their location and their stability allows 
us to estimate the structure of the phase space in their neighborhood and, thus, 
assess their usefulness in supporting observed morphological structures. 

A standard numerical method for finding periodic orbits in a Hamiltonian system 
is the Newton method (see, e.g., \cite{gcobook}, Section 2.4.2). Since we 
considered the bar to be almost along the $y$-axis, it was more convenient to 
apply the Newton method by starting integrating orbits from the $y=0$ axis. Then, 
from Equation~(\ref{eq:hamilton}), we could obtain $\dot{y}_0$, and so the initial 
conditions of a PO can be defined by the coordinates $(x_0,\dot{x_0})$.

The stability of a PO is characterized by the value of its H\'{e}non index 
``$\alpha$'' \cite{h65}. Using H\'{e}non's definition for this index, we 
see that an orbit is stable for $|\alpha| < 1$ and unstable 
otherwise. In the former case, the PO traps around it a set of non-periodic, 
``regular'', orbits reinforcing a morphology according to their topology 
\cite{p05}. In the latter case, the POs repel the orbits in their neighborhood 
and the nearby, ``chaotic'', orbits can be found in a volume of phase space called a 
``chaotic sea''. However, chaotic orbits can be structure-supporting during a 
limited time period as well, if they are ``sticky'', either around an island of 
stability, or close to the unstable asymptotic curves of unstable periodic 
orbits, in Poincar\'{e} cross sections \cite{ch08}. Thus, the areas 
that support a morphological feature, such as a bar, do not overlap, in all cases, 
with the islands of stability we find in the Poincar\'{e} surfaces of a section~
\cite{chpb11}.

Starting with a particular PO and varying continuously a parameter of the 
system, e.g., $E_J$, we find a set of POs that we call a family 
\cite{poinc1892}. 
The knowledge of the basic families of periodic orbits and the evolution of 
their stability as $E_J$ varies gives us the basic, essential, information for 
the global dynamics of a system. A useful tool for following the global 
dynamics of the system is the ``characteristic'' of a family, which is the 
curve that gives the initial conditions $(x_0, \dot{x_0})$ of its POs as a 
function of $E_J$.

\begin{figure}[H]
\includegraphics[width=12 cm]{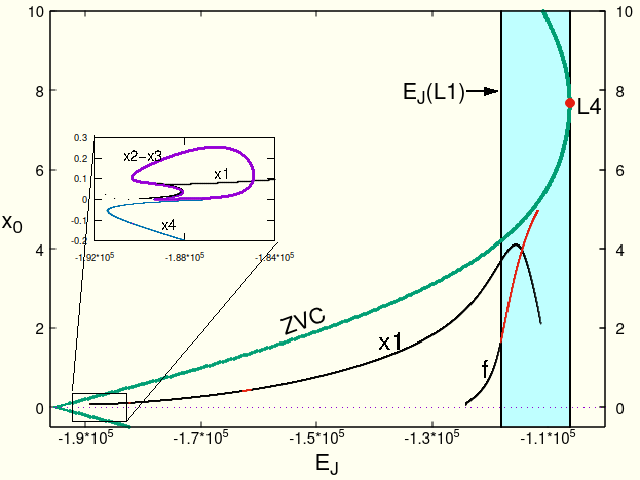}
\caption{Characteristics of the main families of POs, x1 and f, projected 
in the ($E_J$, $x_0$) plane. Stable 
parts of these curves are plotted with black, while unstable with red, color. 
The zero-velocity curve is the thick green curve, having at its local maximum the 
L$_4$ point. The area between the $E_J$'s of L$_1$ and L$_4$ is shaded with 
cyan color. In the embedded frame, we give an enlargement of the 
characteristics of all main families close to the center of the system (see text). The x2-x3 
loop is plotted with magenta, the x1 characteristic with balck and that of x4 
with blue color. 
\label{charact}}
\end{figure}

The characteristics of the families that play a role in the dynamics of the 
system we examined are given in Figure~\ref{charact}. The main families that play 
a role in the dynamics of the model are x1 
and f. The main family is x1, while f is found close to 4:1 resonance. In our 
system, the POs have $\dot{x_0}\neq 0$ and, thus, Figure~\ref{charact} is a 
projection of the $(E_j,x,\dot{x})$ characteristic on the $(E_j,x)$ plane. The 
green curve is the zero velocity curve (ZVC). It separates the regions where 
motion is allowed and we can plot the characteristics of the families from 
those where motion is forbidden. The location of L$_4$ is indicated with a red 
dot in the right part of the figure.

The orbital dynamics close to the center are complicated with more families 
playing locally important roles. Besides x1, we encounter the family of 
retrograde orbits x4, as well as the x2-x3 loop with orbits, in general, 
perpendicular to the bar. However, in our model, they provide, along the x2-x3 
loop, orbits with various inclinations with respect to the 
minor axis of the bar. In the embedded in Figure~\ref{charact} frame, we present 
the projection of the characteristics of these families for $-\text{192,000} < E_J < 
-\text{184,000}$ in the $(x_0, E_J)$ plane. Since Figure~\ref{charact} is a projection, 
there are orbits from different families sharing the same $x_0$ at an $E_J$, 
while they differ in their $\dot{x}_0$ initial condition. Details about the 
orbital dynamics in the very central region of this and other models of the 
general type of Equation~\ref{eq:potrthfour} will be presented elsewhere.

\begin{figure}[H]
\includegraphics[width=13.5cm]{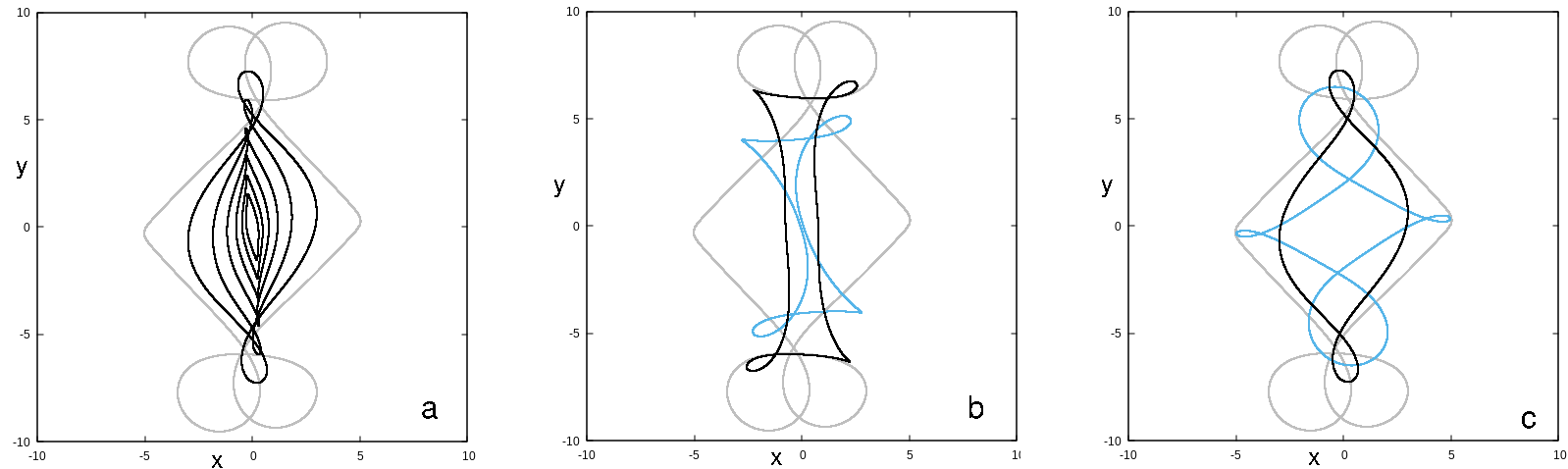}
\caption{Periodic orbits with loops. (\textbf{a}) Stable orbits of x1 are 
plotted with black and an unstable f with pairs of loops at its apocenters is plotted with 
grey. (\textbf{b}) Two stable f orbits (black and cyan) and an unstable in 
grey. (\textbf{c}) POs with loops beyond the 4:1 resonance. The grey belongs to f, while
the two others (black and cyan) to x1.
\label{pos}}
\end{figure}

The important families of POs for the present study are the following:

\begin{itemize}

\item The family x1.
The main building block for the bar of the model is, as expected, the x1 
family. However, we have to underline that, due to the presence of the sine 
terms in Equation~(\ref{eq:potrthfour}), its orbits cross the $y=0$ axis with 
$\dot{x}\neq 0$. We find that essentially all x1 orbits are cuspy and they 
develop loops, which are conspicuous for $E_J > -1.4\times10^5$. 
Representatives of the x1 family are depicted with the black color in 
Figure~\ref{pos}a. All of them are stable, with the orbit with the largest 
loop and the longest projection on the $y$-axis being at $E_J=-$121,468.

\item The family that we indicate as ``f'' in Figure~\ref{charact} has a 
stable and an unstable part. As we observe in Figure~\ref{charact}, it changes its 
stability practically at $E_J(\rm{L}_1)$. Orbits of this family are given in 
Figure~\ref{pos}b. The cyan- and black-colored orbits are stable at $E_J=-$124,157 
and $-$120,000, respectively. The grey orbit, which has developed loops reaching the 
L$_1$ and L$_2$ regions, is at $E_J=-$111,723 and is unstable. As energy 
increases, the $x_0$ initial condition of the orbits of family f increases, 
leading to hexagonal, rhomboidal shapes. We have, in this case, along the 
characteristic of f in Figure~\ref{charact}, a transition from a 4:1 to 6:1 
resonance morphology. Changes in the morphology of POs along a characteristic 
are observed as the curve passes through the region of a resonance (see, e.g.,
Figure 3.7 in~\cite{gcobook}).
For $E_J \gtrapprox -1.21\times10^5$, the POs of family f are unstable. Thus, 
there are no stable, rectangular-like orbits, which could help the bar reach 
corotation. The x1 orbits at these energies already have  big loops at the 
increasing part of the characteristic, for $E_J> -$121,468. They also have 
shorter projections on the $y$-axis than the outermost x1 orbit, drawn in 
black in Figure~\ref{pos}a. The situation with the orbital loops at and beyond the 
4:1 resonance region is summarized in Figure~\ref{pos}c, where we give x1 at 
$E_J=-$121,468 (black), x1 at $E_J=-$112,503 (cyan) and f at 
$E_J=-$111,723 (gray).

\end{itemize}

\subsubsection{Non-Periodic Orbits}
\label{nporbits}

The knowledge of the location and the stability of the periodic orbits provides 
valuable information for the dynamics of a system. Nevertheless, despite the 
fact that POs are the building blocks of morphological structures, their existence is 
practically improbable in real galaxies or even in response models 
such as those we present here. The orbits followed by the particles in 
the response model in Figure~\ref{stellarespo} belong either to some invariant 
curve around a stable periodic orbit in the $(x,\dot{x})$ surface of sections, 
or they travel in the surrounding chaotic seas.

\begin{figure}[H]
\includegraphics[width=13.5cm]{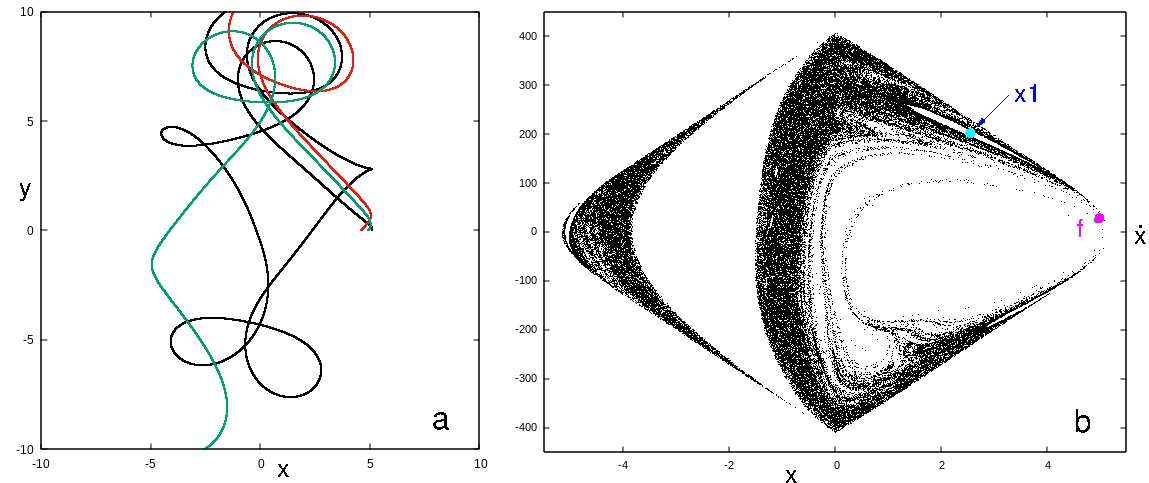}
\caption{(\textbf{a}) Three orbits with loops, plotted with black, 
green and red color) that follow different paths and eventually 
cross the corotation region $((x_{L_{1,2}}, y_{L_{1,2}}) = (\pm1.21, \pm9.65))$. 
(\textbf{b}) The central part of the $(x,\dot{x})$ surface of
section at $E_J=-$111,723. The location of x1 and f are indicated with a green 
and magenta dot, respectively. The left empty part (roughly for $x<-0.5$) is 
occupied by invariant curves around x4 (not plotted), while the empty part for 
$x>-0.5$ is due to the fact that orbits with initial conditions 
in this region are practically escape orbits (they intersect the surface of section 
at large distances, outside the frame of panel).
\label{escsos}}
\end{figure}

For $E_J > -1.21\times10^5$ (Figure~\ref{charact}), there is little space for 
regular motion, because the x1 stability islands are characterized by very thin 
and elongated invariant curves on the surfaces of the section we 
calculated and the f POs are unstable. Nevertheless, we find sticky zones of 
two kinds, namely, (a) around the x1 islands and (b) stickiness in chaos, along 
the unstable asymptotic curves of f \cite{ch08}. Thus, non-periodic orbits at 
these energies will, in most cases, demonstrate a sticky behavior. This means that 
they will imitate quasi-periodic motion for some time and then visit the 
available phase space. All chaotic orbits with loops in the region of the L$_1$ 
and L$_2$ Lagrangian points will eventually cross corotation and follow 
retrograde motion at larger distances, in the frame of 
reference rotating with $\Omega_b$. A large 
percentage of orbits, though, visits the regions close to the 
unstable equilibrium points immediately and cross again the surface of section 
at large $x$, away from the bar. In Figure~\ref{escsos}a, we give three such orbits 
for $E_J=-$111,723, while in Figure~\ref{escsos}b, we present the central part of 
the $(x,\dot{x})$ surface of the section for the same energy composed by the sticky 
orbits of the two kinds we mentioned. The left empty area in Figure~\ref{escsos}b, 
for $x<0$, is occupied by (not drawn) invariant curves around the retrograde PO 
x4. However, the empty region at the right-hand side of the surface of the section, 
roughly for $x>0$, is mainly due to the fact that the orbits with initial 
conditions in this region cross again the surface of the section at large distances, 
i.e., they are practically escape orbits.

\begin{figure}[H]
\includegraphics[width=7.5cm]{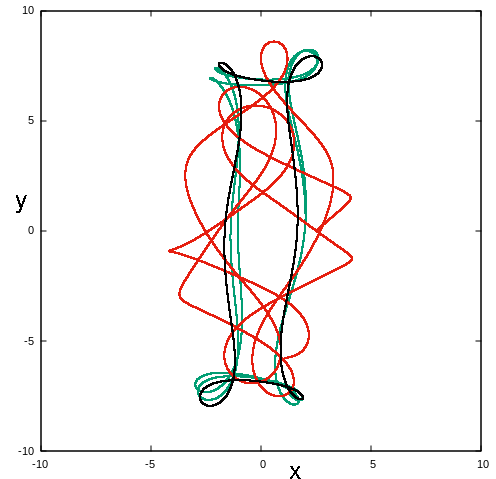}
\caption{The f PO for $E_J=-$118,092 (in black) together with two sticky 
periodic orbits (red and green) at the same energy, integrated for about three 
periods of f. The orbits do not reach the unstable Lagrangian points, located at 
$(x_{L_{1,2}}, y_{L_{1,2}}) = (\pm1.21, \pm9.65)$.
\label{stick}}
\end{figure}

Orbits sticky to the small stability islands of x1 and f for $E_J < -\text{11,808} = 
E_J(\rm{L_1})$, do not cross corotation. In addition, sticky to chaos orbits 
associated with the unstable f orbits close beyond the transition of the 
family to instability, stay for long periods inside corotation. The morphology of such 
orbits is presented in Figure~\ref{stick}, which gives the f PO for $E_J=-\text{118,092}$ 
(in black) and two more sticky orbits (in green and red), integrated for 
about three periods of f. At this energy, the f family has just become unstable 
and is very close to $E_J(\rm L_{1,2})=-\text{118,080}$. The sticky orbits for $E_J 
< E_J(\rm L_1)$ build the component that we call the bar's ``envelope'' 
(Figure~\ref{stellarespo}g,h), which surrounds the x1 bar.


\section{Gas Response}
\label{seq:gresponses}

We studied the response of the gas to the barred potential 
of Equation~(\ref{eq:potrthfour}) by means of two hydrocodes, namely, the Lagrangian 
scheme smoothed particle hydrodynamics (SPH) \cite{gm77, l77} and the adaptive 
mesh RAMSES code \cite{ramses}. We did this in order to compare the predicted 
flows by means of the two codes and find possible differences due to the 
different derivations of the equations of hydrodynamics in the two cases.

The potential is introduced in the gas models in the same way as we described in 
Section~\ref{sresponses} for the stellar response models. We started again from a 
10~kpc disk homogeneously populated with particles in a circular motion in the 
axisymmetric part of the potential, $\Phi_0 (r)$ and we linearly increased the 
bar term within three pattern rotations until its full value. The gas is not 
self-gravitating.

\subsection{SPH}
\label{sph}

For the SPH calculations, we used the code described in \cite{pa00}, while test 
runs were performed with the SPH codes used in \cite{bbp95,kw02}, 
obtaining practically the same results. Starting with a disk with $4\times 10^4$ 
SPH particles that play the role of
tracers in order to follow the gas properties by means of the Lagrangian code, 
we ran a number of isothermal models with sound speeds, $c_s$ = 5, 10, 20 
and 30~km s$^{-1}$, combined with artificial viscosity parameters $(\alpha, 
\beta) = (0.5, 1)$ and (1, 2).

\begin{figure}[H]
\includegraphics[width=13.5cm]{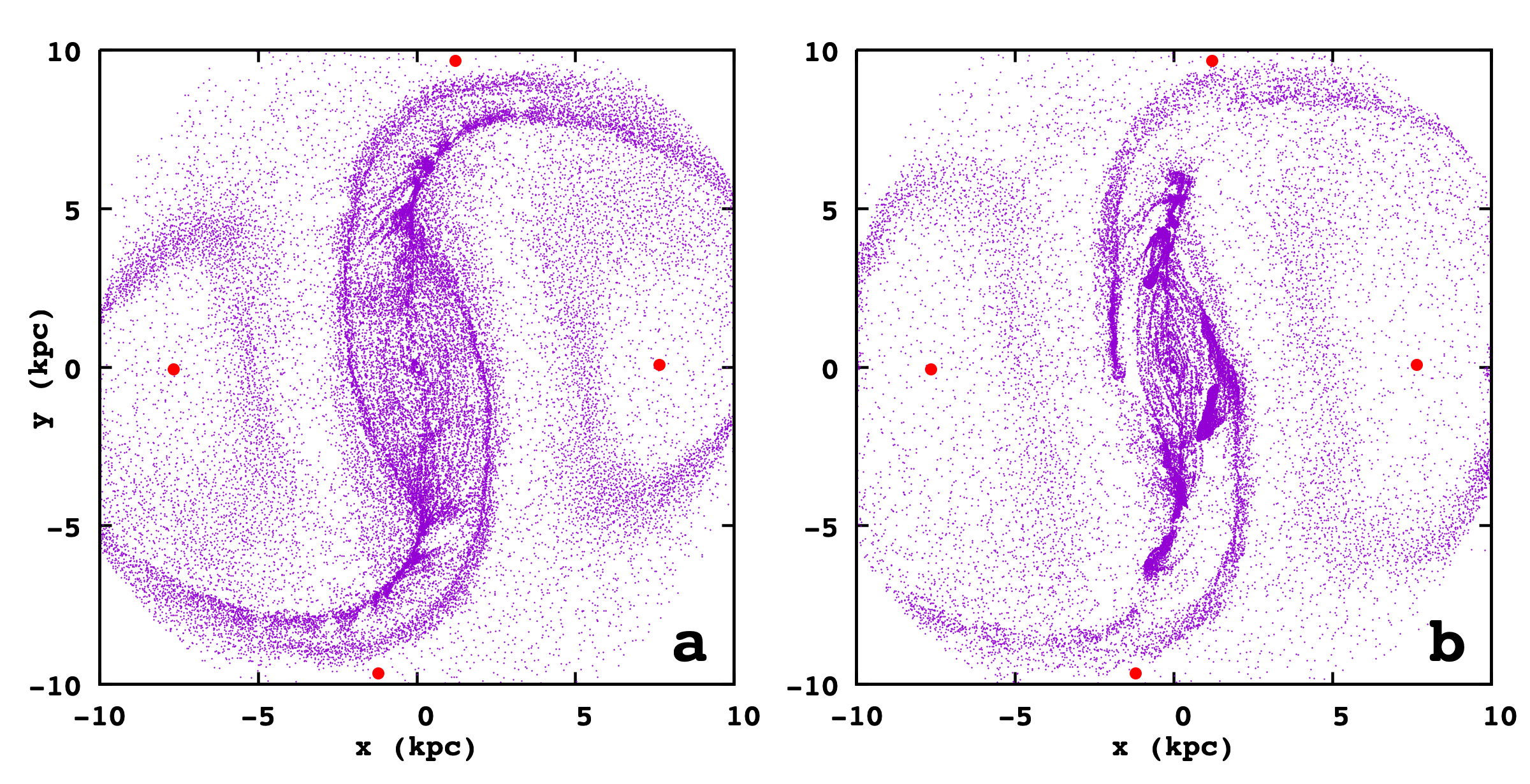}
\caption{The SPH response without redistribution of the particles, during the 
growing phase of the bar, for $t=0.52$ (\textbf{a}) and for $t=0.65$ (\textbf{b}). The artificial 
viscosity parameters we used are $(\alpha, \beta) = (0.5, 1)$. In this, and in 
all relevant subsequent figures, the bar rotates counterclockwise. Red dots 
indicate the Lagrangian points of the model.
\label{sphmod}}
\end{figure}

In all these models, we had a smooth response during the major part of the 
growing phase of the bar. However, for times close to three pattern rotations, there 
were already clumps formed in the high-density regions of the disk. On top of 
this, a considerable percentage of gaseous particles was crossing the corotation 
region without practically returning back to it at later times. The overall 
morphology during the growing phase was as in Figure~\ref{sphmod}a, while after 
almost three pattern rotations, as in Figure~\ref{sphmod}b. In the model of 
Figure~\ref{sphmod}, we have $c_s$ = 10~km s$^{-1}$ and $(\alpha, 
\beta) = (0.5, 1)$. This is qualitatively in 
agreement with the responses found in \cite{lsh98} (see, e.g., their Figures 2 or 3). 
This is a characteristic response of this potential, independently of the 
hydrodynamical parameters of each individual model. We note that our 
calculations are in the frame of reference that rotates with $\Omega_{b} 
=27$~km~s$^{-1}$~kpc$^{-1}$.

As time increases, the number of gaseous SPH particles is reduced. This happens 
because a part of them is trapped in clumps of irregular dense regions, not 
observed in barred galaxies in the local universe (except in some late-type 
galaxies) and because another part of 
them crosses the corotation region and practically escapes. We also observe that 
the regions around L$_4$ and L$_5$ become gradually empty. As a result, it 
becomes problematic to follow for a long time the evolution of the shocks and 
the overall morphology of the gas features in the disk. The situation improves 
by reducing the artificial viscosity, i.e., by using $(\alpha, \beta) = (0.5, 1)$ 
instead of (1,2) and by increasing the sound speed of the models. Nevertheless, 
the improvements by varying the parameters in the response models are small. 
Replenishment of the ``lost'' particles is necessary in order to follow the 
evolution of the model for later times.

\textls[-15]{Replenishment of the ``escaping'' particles was initially applied by 
removing particles reaching distances larger than 11--12\,{kpc} from the center 
and relocating them at random positions within a disk with a 10~kpc radius. 
They were given velocities corresponding to a circular motion in $\Phi_0$. Such a 
modification in the code was already used in the model we 
presented in Figure~\ref{sphmod}. However, as we can observe, the main problem 
remains in the over-dense regions, since these conglomerations are not dissolved 
with time, practically preventing the smooth formation of the dust-lane shocks 
in the bar region. Thus, an additional correction was necessary for preventing 
the trapping of particles in the spurious overdense regions. For this purpose, 
we considered an additional grid and a threshold of the numerical density of 
particles. In the various models, we tried grids ranging from $100 \times 100$ 
to $500 \times 500$ and limiting numerical densities ranging from 5 to 50 
particles/cell. When the critical numerical density in a cell was exceeded, each 
particle had a 50\% probability to be displaced in a random place in the disk, 
following the recipe for the ``escaping'' particles. This way, we avoided the 
formation of the irregular clumps, while we did not introduce into the model empty 
holes, which would have appeared if all particles of the clump were to be 
relocated.}

\begin{figure}[H]
\includegraphics[width=9.75cm]{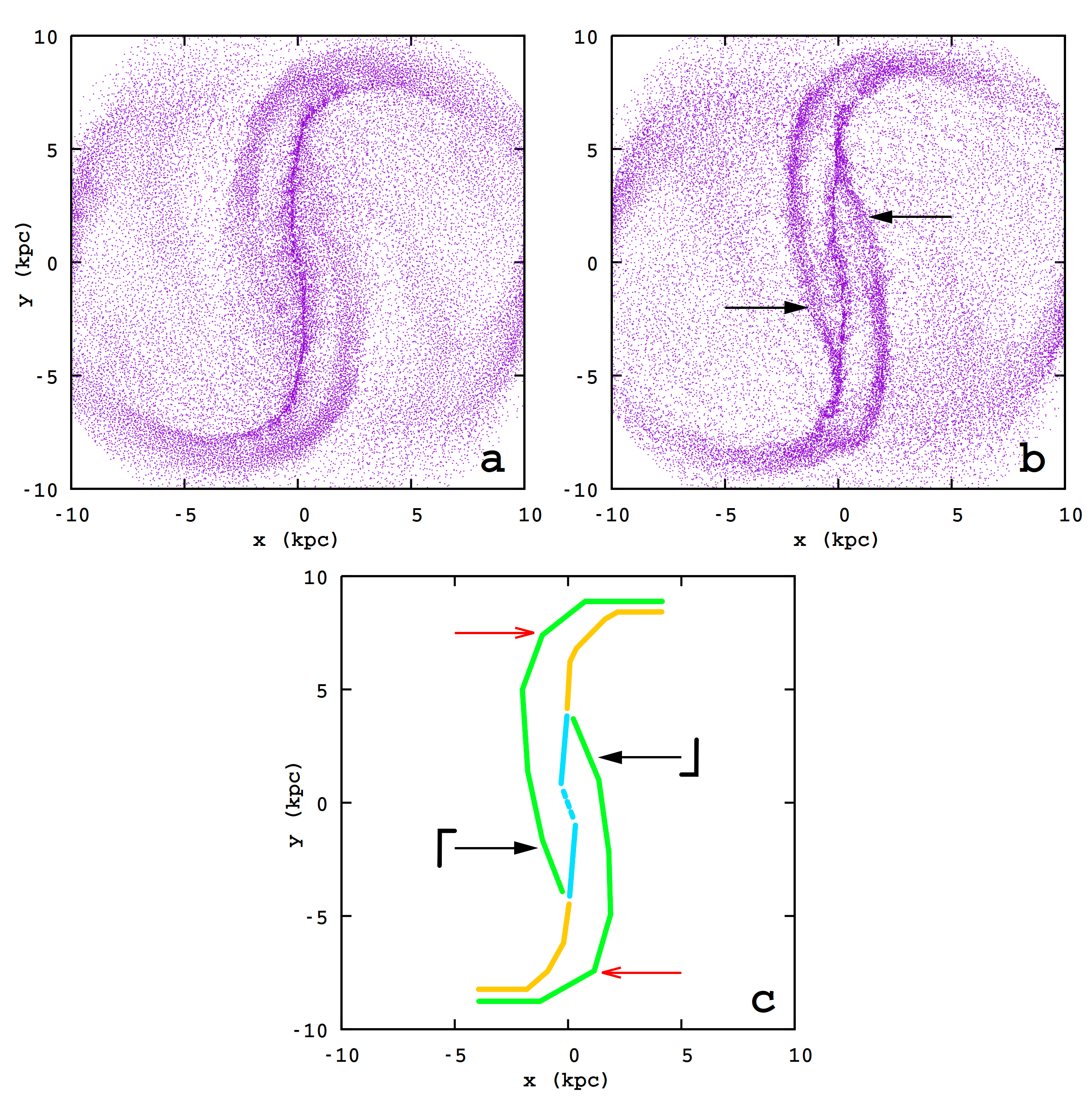}
\caption{Characteristic snapshots of a typical model with redistribution of 
particles in action, at $t=0.53$ (\textbf{a}) and $t=9.83$ (\textbf{b}). In (\textbf{c}) we give a sketch 
that outlines the dense parts of the model. In the text, the green features, 
indicated with arrows, are described as ``$\Gamma$/\reflectbox{L} tails''. We 
draw the straight-line dust lane shocks and its continuations with cyan and 
yellow color respectively. 
\label{sphrepl30}}
\end{figure}

\textls[-15]{The response models could be followed then for a time as long as needed, 
establishing a rather invariant response morphology, as the one depicted in 
Figure~\ref{sphrepl30}, which is a typical case. In the model presented in 
Figure~\ref{sphrepl30}, we have $c_s=20$~km s$^{-1}$, $(\alpha,\beta) = (0.5, 1)$, 
$r_{max}=12$\,kpc, a local grid $100 \times 100$ and a critical numerical 
density of 40 particles/cell, for applying the redistribution of the particles. In 
this case, for times larger than the time when the potential has reached its final 
value, about 20,000 particles cross the 12~kpc radius during one bar rotation 
period. We stopped integrating these particles, since, in any case, the potential we 
used is not accurate at those distances, and we replenished the disk with an equal 
number of particles at random positions. In addition, in order to avoid the formation of 
clumps, in average 3.4 particles per step were displaced away from the 
over-dense regions. Let us also note that these clumps tended to be formed always 
at roughly the same locations.}

Straight-line dust-lane shocks are typical of the response models, suggesting a 
strong bar \cite{a92b}. In our model, they are found slightly displaced ahead of 
the major axis of the bar, reaching $y\approx 4.5$\,kpc. Beyond that point, they 
continue existing at larger distances from the center, turning eventually 
towards the direction opposite to the rotation of the system. In the two cases 
of the SPH models we presented in Figures~\ref{sphmod} and \ref{sphrepl30}, the 
flows are qualitatively similar. In addition to the straight-line shocks, they are also 
characterized by the presence of elongated filament-like structures, which appear 
bifurcating from the straight line shocks and surround them. They are indicated 
with black arrows in Figure~\ref{sphrepl30}b. An outline of the dense regions of 
the response model is given in the sketch of Figure~\ref{sphrepl30}c. The loci of 
the inner part of the straight-line shocks is given in cyan and those of the 
outer one in yellow color. The filament-like structures are drawn in green. Due 
to their characteristic turn towards the trailing side of the bar, roughly 
beyond the points indicated with red arrows in Figure~\ref{sphrepl30}c, we 
describe them as having a ``$\Gamma$-like'' morphology (``\reflectbox{L}'' for 
the one associated to the lower part of the bar).

\begin{figure}[H]
\includegraphics[width=10.5cm]{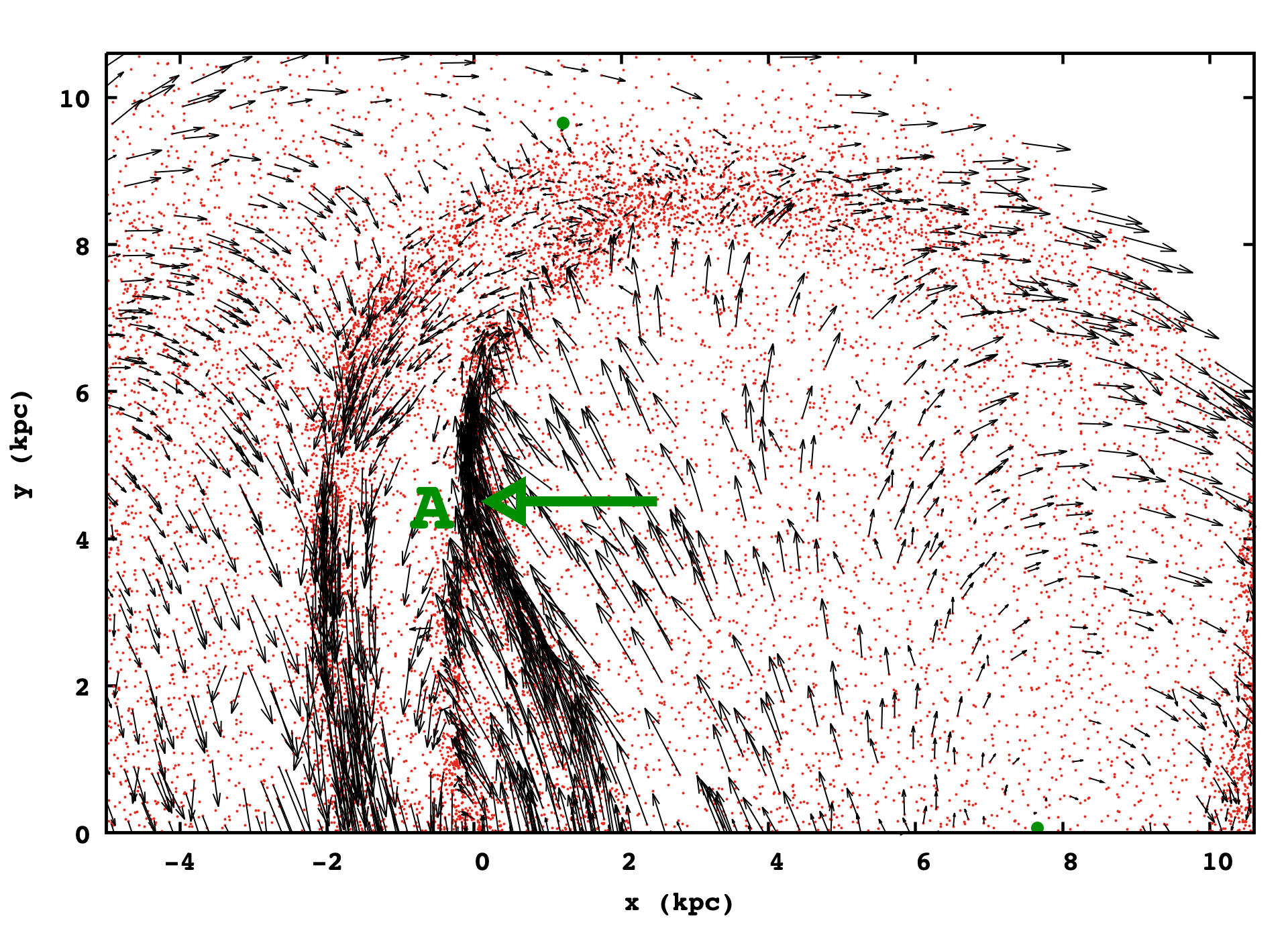}
\caption{The velocity field of the snapshot of Figure~\ref{sphrepl30} is given by 
zooming into its $(-5,11)\times(0,11)$ region. The green arrow indicates point A, 
where the tail of the lower $\Gamma$ feature joins the straight-line shocks. 
The green dot at $(x,y)=(1.21, 9.65)$ indicates the location of L$_1$.
\label{vf1sph}}
\end{figure}

In Figure~\ref{vf1sph}, we present the velocity field, always at the rotating with 
$\Omega_b$ frame of reference, as it was finally developed in the snapshot 
of Figure~\ref{sphrepl30}b. Similar velocity fields were found in 
\cite{a92b}, suggesting that they are generic. Zooming into 
this particular region, for better understanding the details, we can 
follow the flow all over the disk. The critical point is A at $(x,y)\approx 
(0,5)$, at the location where the green arrow points. The straight line shocks 
are formed below A, slightly to the left of it. Gas is flowing in the direction 
of rotation, upwards, in the part of the model we plot in Figure~\ref{vf1sph}, is 
shocked and then flows down-streams ahead of the straight-line shocks. The 
other branch that reaches A (below and to the right of it) belongs to the 
$\Gamma$ feature that reaches the L$_2$ region, i.e., an ``\reflectbox{L}'' 
structure (not depicted in Figure~\ref{vf1sph}). It is the right ``bifurcating'' 
branch indicated with an arrow in Figure~\ref{sphrepl30}. We observe that the gas 
is flowing towards A. At the left of A, at $x\approx -1.8$\,kpc, we can see the 
origin of the corresponding feature reaching the bifurcating point at the lower 
part of the model. Gas is streaming in this region like in a funnel. We will 
discuss the flow in the $\Gamma$ feature again, in comparison with what we found 
by means of the Eulerian RAMSES scheme in Section \ref{ramses}. For the time 
being, we note that, according to Figure~\ref{vf1sph}, the flow in the upper part of 
$\Gamma$ is split in two parts, one in the direction of rotation and another one 
in the opposite direction. The former is characterized by low velocities, which 
nevertheless lead the gas in the funnel of the tail of the $\Gamma$ feature. The 
latter is associated with the flow around L$_4$ and can be traced, because of 
the continuous replenishment of the ``lost'' particles in the model. As we can 
see in the left part of Figure~\ref{vf1sph}, the tail of $\Gamma$ is also reinforced 
 by the flow of particles around L$_5$ (not included in the figure).
\vspace{-16pt}

\begin{figure}[H]
\includegraphics[width=10.5cm]{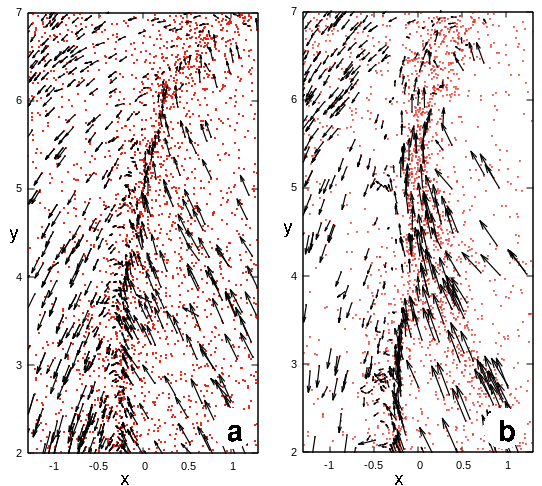}
\caption{The flow in the straight-line shocks region in two typical cases. 
(\textbf{a}) During the growing bar phase. (\textbf{b}) During the time the bar has a constant 
amplitude $(t> 3T)$. We can observe in (\textbf{b}) how the particles of the bifurcating 
tail help the shock extend to larger distances from the center, as they flow 
upwards in the direction opposite to the rotation of the system.
\label{vf2sph}}
\end{figure}

\begin{figure}[H]
\includegraphics[width=9.0cm]{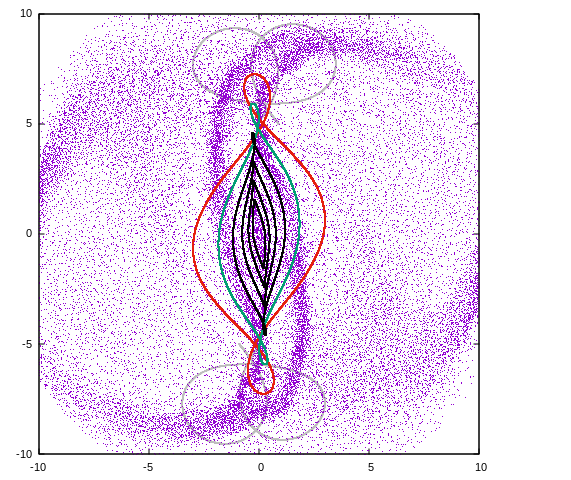}
\caption{Stable x1 orbits (black, green and red) and the (grey) loops of 
an 6:1 f orbit (see Figure~\ref{pos}). The straight-line shocks are formed in the region 
occupied by the black x1 POs. The shocks avoid the loops, while the unstable f 
orbit does not affect the flow.
\label{orbsph}}
\end{figure}

Finally, we focused on the flow of gas in the straight-line shocks region, since 
this is the most typical feature of hydrodynamical models in bars \cite{a92b}. 
We find a small but evident qualitative difference in their morphology comparing 
snapshots during the initial growing bar phase with snapshots during the phase 
with the constant amplitude of the potential. This can be understood by 
following the gas flow in the two cases and is given in Figure~\ref{vf2sph}. 
During the growing bar phase, the straight-line shocks are longer, with 
$y\lessapprox\,6$\,kpc, as can be seen in Figure~\ref{vf2sph}a. The shocks are 
formed in the standard way \cite{a92b}, namely, as they move up stream in the 
direction of rotation, they are shocked along the straight-lines, loose velocity, 
and ahead of the shock the flow continues down stream. During the constant bar 
amplitude phase (Figure~\ref{vf2sph}b), the straight lines hardly reach the 
``bifurcating'' point at $y\approx 4.5$\,kpc. At larger heights, they are 
gradually replaced by the particles that form the tails of the $\Gamma$ feature. 
Their motion is opposite to the rotation of the system. Considering just the 
morphology of the shocks, one cannot discern the qualitative changes in the flow 
along the shocks. In Figure~\ref{sphrepl30}b, for example, the shocks close to the 
major axis of the bar appear along a unique curve reaching $y\lessapprox\,6$\, 
kpc. Only by careful inspection can we  realize that, just beyond the point at 
which the tails of the bifurcating $\Gamma$/`\reflectbox{L}' features appear, 
the shocks change inclination, becoming more perpendicular to the $x$-axis. In 
Figure~\ref{vf2sph}b, we realize that this ``perpendicular part'' has $4 < y < 
6$\,kpc, where we can observe velocity vectors parallel to the $y$ axis.
The question that arises is what is the orbital background behind the flows 
described in the two last figures. In Figure~\ref{orbsph}, we plot a set of orbits 
over the snapshot of the SPH response model of Figure~\ref{sphrepl30}b. The 
black, green and red orbits belong to x1 and are stable. The straight-line 
shocks appear at the leading sides of the black x1 POs. The green one has 
discernible loops that appear ahead of the dense shock lanes. The red PO is 
the longest x1 orbit with loops, since, for larger $E_J$s, the orbits of the 
family shrink in the $y$ and expand in the $x$ direction (Figure~\ref{pos}c) 
and it is not evident from Figure~\ref{stellarespo} that they are populated at all. 
The shocks are arranged in a such a way as to leave the major part of the 
loops ahead of them. On the left side of the loops, we observe that the 
$\Gamma$ feature also avoids  the region of the loops. Finally, we plot with 
gray just the loops of the unstable f 
orbit depicted in Figure~\ref{pos}. It does not seem to be directly associated 
with the flow in the region, as it does not seem to affect the stellar flow in
Figure~\ref{stellarespo}.


\subsection{RAMSES}
\label{ramses}

In order to compare the gas responses to our potential in Lagrangian and 
Eulerian schemes, we used, for the latter, as earlier mentioned, the 
publicly available code RAMSES\endnote{available in 
\url{https://bitbucket.org/rteyssie/ramses/src/master/}, accessed on 
16 May 2022} \cite{ramses}. It has 
been successfully used in gas dynamical studies of barred and non-barred spiral 
galaxies \cite{fdo16,ff16}. Gas dynamics in RAMSES is computed with a 
second-order unsplit Godunov scheme. It is an adaptive mesh refinement scheme 
an,d in the models we present in our study, the maximum resolution used reaches 
30~pc. Models with a maximum resolution of 60~pc were also used. In all cases, the 
gas was considered isothermal with an adiabatic index 5/3. We ran 
several models with sound speeds $c_s=5, 10, 20$ and 30~km s$^{-1}$, as in SPH.

For test purposes, we carried out a complete study with the potential in 
\cite{a92b}, using all Ferrers models of that work having index n = 1, with the 
RAMSES code \cite{pst22}. All responses gave identical results to the 
corresponding models in \cite{a92b} as regards the morphology and a very similar 
description of the dust-lane shocks. This is important, since it confirms that any 
differences between the morphology of the gas flow in \cite{a92b} and here will 
be due to the different potential and not to inadequacies of one code or another.

\begin{figure}[H]
\includegraphics[width=10cm]{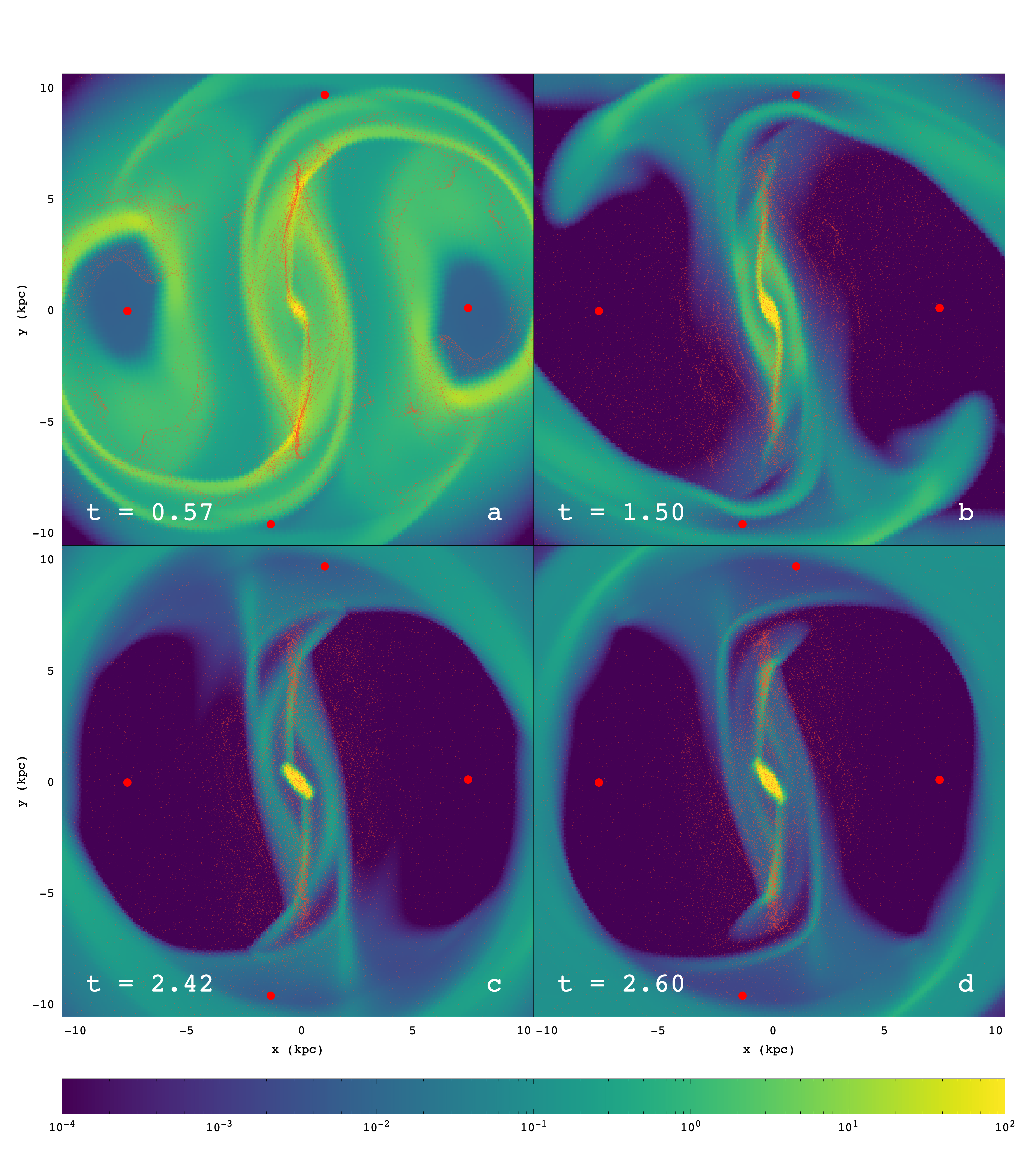}
\caption{The evolution of a typical RAMSES model, with $c_s=5$\,km s\,$^{-1}$. The snapshot 
in (\textbf{a}) is in the growing bar phase, while in (\textbf{b}--\textbf{d}) the bar 
perturbation has already reached its maximum. Times are given in the lower-left 
corner of the frames. Densities increase from left to right according to the 
logarithmic scale of the color bar at the bottom of the figure. In red, we plot 
the stellar response particles at the corresponding times. The bar rotates 
counterclockwise. Red dots indicate the location of the Lagrangian points.
\label{ramses5evol}}
\end{figure}

In all RAMSES models, we find again a typical overall response, where the main 
morphological features, and the corresponding flows, are the same as in SPH. 
Nevertheless, the details of the responses are different when different sets of 
the free parameters (sound speed, grid resolution, etc.) are used. A typical 
response is given in Figure~\ref{ramses5evol}, where the evolution of a model with 
$c_s=5$\,km s\,$^{-1}$ and maximum resolution of 60~pc is presented. We start 
again with initial conditions homogeneously populating a 10 \,kpc disk and 
velocities corresponding to the circular motion in the axisymmetric part of 
the potential in Equation~(\ref{eq:potrthfour}). As in the stellar and SPH responses, the 
potential grows to its full strength over three pattern rotations (we remind 
that $\Omega_b = 27$\, km s$^{-1}$ kpc$^{-1}$; therefore, in our time units, the rotational 
period is $T=0.23$). Thus, the flows in all cases can be compared. In addition, in this 
section, all figures are in the frame of reference rotating with $\Omega_b$.

In Figure~\ref{ramses5evol} we give four typical snapshots. Times appear in the 
lower-left corner of the panels. Density increases according to the color bar are at 
the bottom of the figure from left (dark blue, minimum) to right (yellow, 
maximum). The snapshot in Figure~\ref{ramses5evol}a is after about two pattern 
rotations, i.e., still in the growing phase of the bar. The dust-lane shocks are 
already discernible, as well as the branches joining them at about $5\,kpc$. 
Towards the ends of the bar, they continue as spiral arcs with a double character. This 
set of double spiral arms extends mainly inside corotation (we remind that 
$(x,y)_{\mathrm{L}_1}$ = (1.21, 9.65) and $(x,y)_{\mathrm{L}_4}$ = (7.657, 0.064)). 
We observe that, already at this phase, we have around L$_{4,5}$ low-density 
regions. For $t\geqq3 T$, these regions are found to be almost (but not completely) empty.
\vspace{-12pt}

For times larger than $3T$ (Figure~\ref{ramses5evol}b--d), when the bar 
perturbation has reached its maximum value, we observe that the straight-line, 
dust-lane shocks, reaching almost $y=5$\,kpc, appear more pronounced than in 
Figure~\ref{ramses5evol}a, while the spiral arcs that appear in the upper-right 
and lower-left quadrants of the snapshots are fainter than during the growing 
bar phase. We also observe that they do not always appear as a continuation of 
the dust-lane shocks, as in Figure~\ref{ramses5evol}a. Gradually, they are 
transferred to $\Gamma$-like features, similar to those developed in those SPH 
models at the corresponding regions of the disk (Figure~\ref{sphrepl30}). We note 
that there is a period, just after the time the bar perturbation reaches its 
maximum value, lasting for about $t\approx1.5\,T$, during which the model is 
characterized by turbulence and the formation of clumps. This is similar to the SPH 
models, in which we do not dissolve the over-dense regions (Figure~\ref{sphmod}b). 
However, in the RAMSES simulations, the clumps that formed conglomerations deform 
with time and gradually dissolve, without reducing their density manually. 
Then, the model reaches a semi-stationary phase, lasting to the end of the 
simulation (after at least 10 pattern rotations). The snapshots in panels 
Figure~\ref{ramses5evol}b-d are typical of the response we find.

In all panels of Figure~\ref{ramses5evol}, the gas is plotted on top of the 
stellar response (drawn in the red color) at the corresponding times. Taking into 
account Figure~\ref{stellarespo} and~\ref{pos}, we realize that the stellar 
response in the regions of the loops is characterized by the presence of 
small, ansae-like features---which are formed ahead, in the direction 
of rotation---of the dust-lane shocks and/or their extensions.

\begin{figure}[H]
\includegraphics[width=13.5cm]{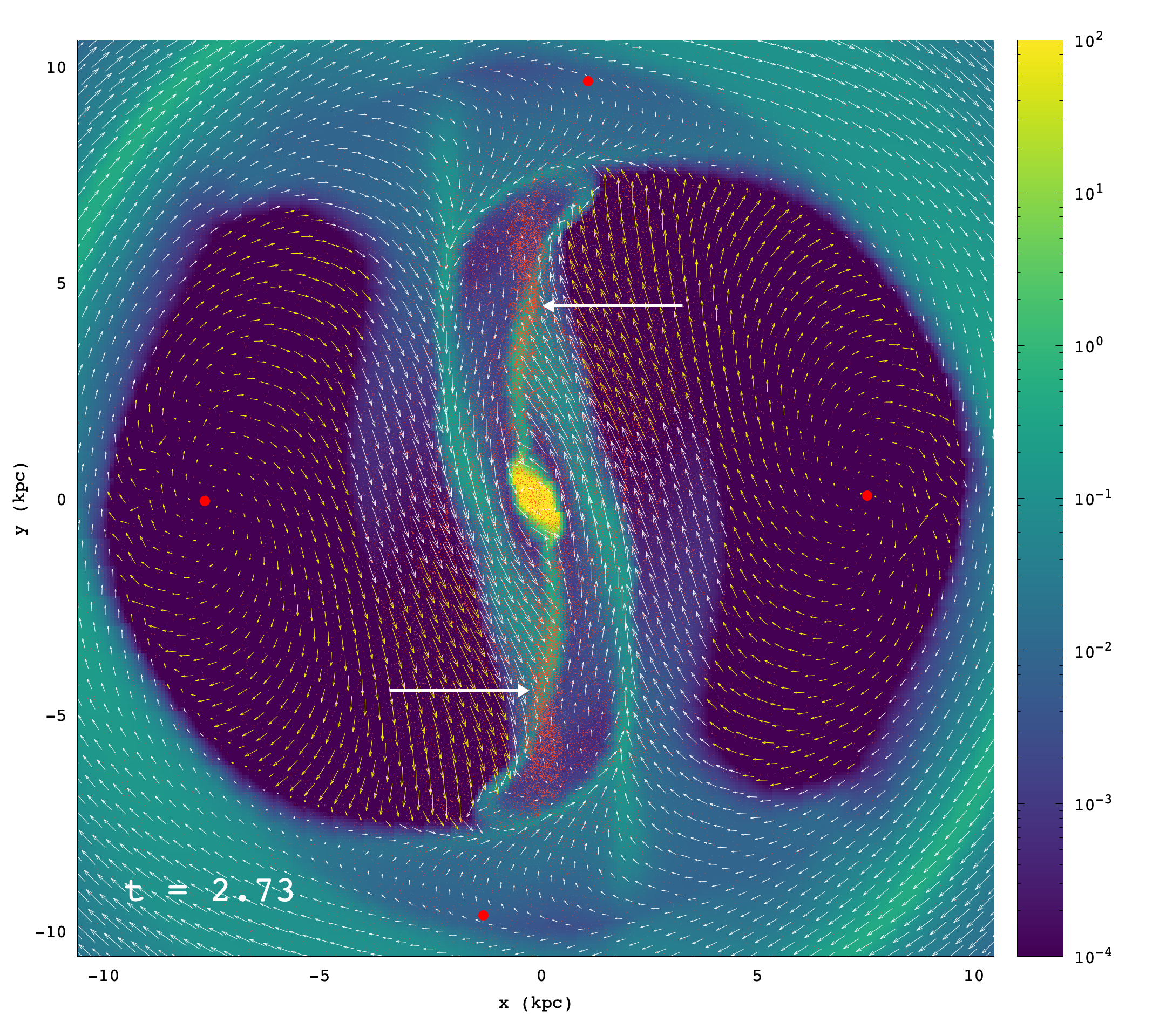}
\caption{A typical snapshot of our RAMSES simulation, summarizing the main 
morphological features, together with the velocity vectors indicating the flow. 
The flow vectors are colored yellow, when they are in 
regions below the minimum density we considered. The two long white arrows 
indicate the point where the tails of the $\Gamma$ features join the dust-lane 
shocks in the bar region. The stellar response is given in the background using a
red color for the test particles.
\label{typicalss}}
\end{figure}

Strictly speaking, the gas morphology in the snapshots of the RAMSES responses 
varies with time, retaining, however, as common features the straight-line 
dust-lane shock loci up to $y\approx4.5$\,kpc, the extensions bifurcating from 
these points, the rather-empty regions around L$_{4}$ and L$_{5}$, as well as a 
x2 region inclined with respect to the $x$-axis x2 region. The last two features 
were also clear in \cite{a92b}. These are summarized in Figure~\ref{typicalss}, 
which we consider as a typical RAMSES response. Breaks of the straight-line 
shocks are also sometimes observed, but last for very small time intervals and 
are reorganized again, so we may consider them as transient features. The 
critical points along the dust-lane shocks at which we observe the bifurcating 
extensions are indicated with big white arrows.
\vspace{-4pt}

In Figure~\ref{typicalss}, we also present the overall flow of the gas, by means of 
the corresponding velocity arrows. Yellow arrows are used for densities below 
the minimum value we include in the figure, as indicated by the color bar at the 
right-hand side of it. Evidently, the flow in the region of the tails is toward 
the points where they join the shock loci. The low gas surface density regions 
(blue) are typical of this, as well as of all other snapshots and several 
previous simulation results \cite{a92b}. They form one on either side of the 
bar. The dense parts of the shock loci, i.e., the inner straight-line parts and 
their curved continuations in the direction opposite to the one of the rotation 
of the system, avoid visiting these regions, which are occupied by 
quasi-periodic stellar orbits trapped around x1 POs with big loops.
To show this, we give in the background of the gas in Figure~\ref{typicalss}, the 
stellar response model of the same time ($t=2.73$). The stellar particles are 
plotted with a red color. The conspicuous reddish ``bulb-like'' features at the 
low-density gas regions, ahead of the dust-lane shocks, close to the tips of the 
long, white arrows, with $y\gtrapprox 4.5$\,kpc and $y \lessapprox -4.5$\,kpc, 
are typical configurations built by quasi-periodic orbits associated with x1. 
We observe only low-density gas in transient passings through these areas, 
without the formation of shocks.

The variation of the morphologies in the RAMSES models refers mainly to the 
extensions of the dust-lane shocks from their straight-line part towards the 
$\Gamma$-like features (their part with $|y|>5$\,kpc in Figure~\ref{typicalss}) 
and the width of the bifurcations of the dust-lane shocks. We find that there is an 
approximate repeating cycle of the flow associated with the destruction and 
regeneration of these extensions, which also affects  the rest of the gas 
morphology. However, during the evolution of the model up to at least 10 pattern 
rotations, each cycle did not have the same period as the others. In addition, during each 
cycle, the morphological evolution was not exactly the same. In all cases it 
lasted always less than $T$. We describe such a cycle in Figure~\ref{ram_stargas}, 
by means of four panels, during a time interval $\Delta t = 0.11$.
In Figure~\ref{ram_stargas}a the gas streaming along the bifurcating point ``A'', 
at $y\approx 5$~\,kpc, indicated by a white arrow pointing to the left, bends 
upwards to the right, meeting the upper branch of the $\Gamma$ structure. The 
arrows depicting the flow point out that the gas streaming towards ``A'' 
bypasses the triangular region occupied by the big orbital loops of the test 
particles, colored red. The flow along the extension is from ``A'' to the upper 
branch of the $\Gamma$ structure. Then, the gas continues flowing along the 
denser, green regions counterclockwise, forms a loop around the rather depleted 
from gas triangular region and joins the flow into the ``funnel'' that forms the 
bifurcating ``tail'' on the left side of the bar, i.e., the side of L$_5$. This 
tail meets the dust-lane shock at the lower (with $y<0$) side of the bar 
(Figure~\ref{typicalss}). Another (heavy yellow) arrow in Figure~\ref{ram_stargas}a 
points to the location on the upper branch of ``$\Gamma$'', at which the flow 
``splits'', pointing to two opposite directions. We remind that all these 
features are inside corotation. The flow along the ``chaotic'' spirals is above 
and to the right of the heavy, yellow arrow and is associated with the clockwise 
pointing arrows of the flow.

In Figure~\ref{ram_stargas}b, the streaming along the bifurcating tail towards 
``A'' is accumulated at the upper end of the dust-lane shock, forming a 
characteristic bending of it to the right. The extension above this point 
towards the upper branch of the $\Gamma$ feature is not observed.

The flow of the gas, as given by the arrows, indicates a weak extension, not discernible at 
the level of densities chosen for presenting the flow. Low-density gas, now 
from a broader area, is streaming into the region ahead of the dust-lane 
shocks. The upper branch of the $\Gamma$ is shifted to larger distances from the 
center and the point along it, at which the flow ``splits'', has moved to the 
left, compared with Figure~\ref{ram_stargas}a (again indicated with a heavy, 
yellow arrow). At a certain time, the accumulated gas conglomeration at the end
of the x1-dust -lane shock recedes, as gas along the bifurcating tail continues streaming 
towards ``A''. Then, the accumulated gas starts flowing upwards, bypassing the 
region where the underlying test particle orbits form loops 
(Figure~\ref{ram_stargas}c). Finally, in Figure~\ref{ram_stargas}d, we reach a 
situation similar to the one encountered in Figure~\ref{ram_stargas}a, with the 
gas flowing around the triangular, low gas density region, occupied by the 
orbital loops.

As we mentioned before, the cycles are not identical and, while being 
relatively short-lived, do not evidently have a specific period. In order to 
find a mean response for our RAMSES model, we considered 100 
snapshots for the period $2 \leqq t < 3$, i.e., during the time we have a quasi-stationary response. 
This period corresponds to 4.35 rotational periods of the system. We stacked 
the images of all these snapshots and the result can be seen in Figure~\ref{mean}.
The result resembles the typical response we presented in Figure~\ref{typicalss}, 
both as regards the average morphology as well as the average flow, which is 
given in enlargement in the upper-right corner of the figure.

The overall response does not change as the sound speed, $c_s$, varies. However, 
there are several differences in the details. We followed the responses of the 
same initial conditions in models with the same resolution but with different 
$c_s$. In Figure~\ref{ramsescs}a, we give the response in a model with $c_s=10$, 
while in Figure~\ref{ramsescs}b, a model with 20~km~s$^{-1}$. There are 
no major differences between the model in Figure~\ref{ramsescs}a and the model 
with $c_s=5$~km~s$^{-1}$ in Figure~\ref{ramses5evol}. However, as $c_s$ 
increases, differences in the models come into view: more conspicuous are the 
instabilities that appear along the extensions towards the upper part of the 
$\Gamma$ structures, i.e., in regions with strong density gradients. In addition, 
the contrast between low- and high-density regions decreases. For example, in the 
regions around L$_4$ and L$_5$ in Figure~\ref{ramsescs}b, we encounter gas 
diffusing, increasing this way the gas density compared with the 
snapshots of models with lower sound speed at the same time. For even larger 
$c_s$'s, these instabilities are more evident.

\begin{figure}[H]
\includegraphics[width=13.5cm]{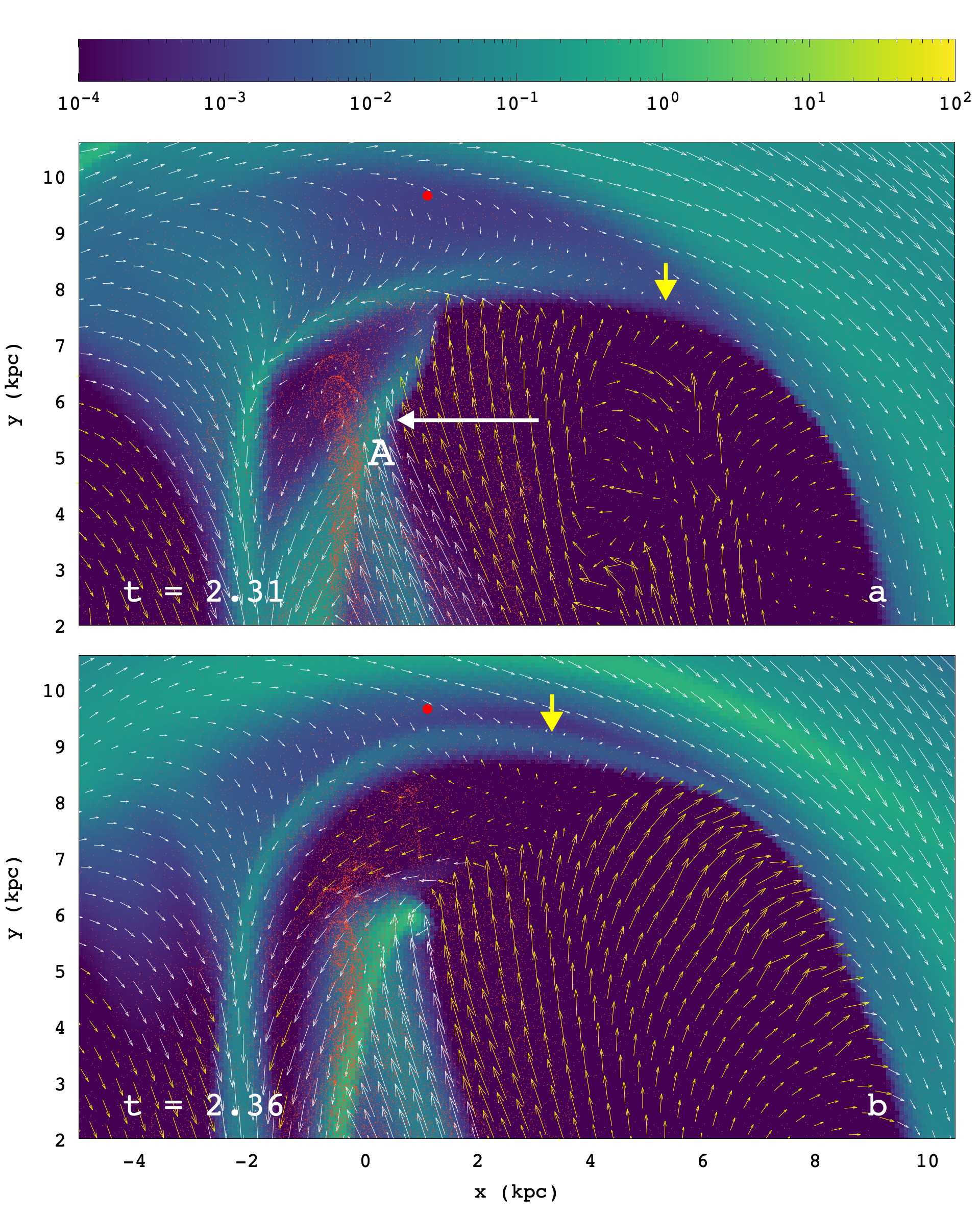}
\caption{\textit{Cont}.}
\end{figure}
\begin{figure}[H]\ContinuedFloat
\includegraphics[width=13.5cm]{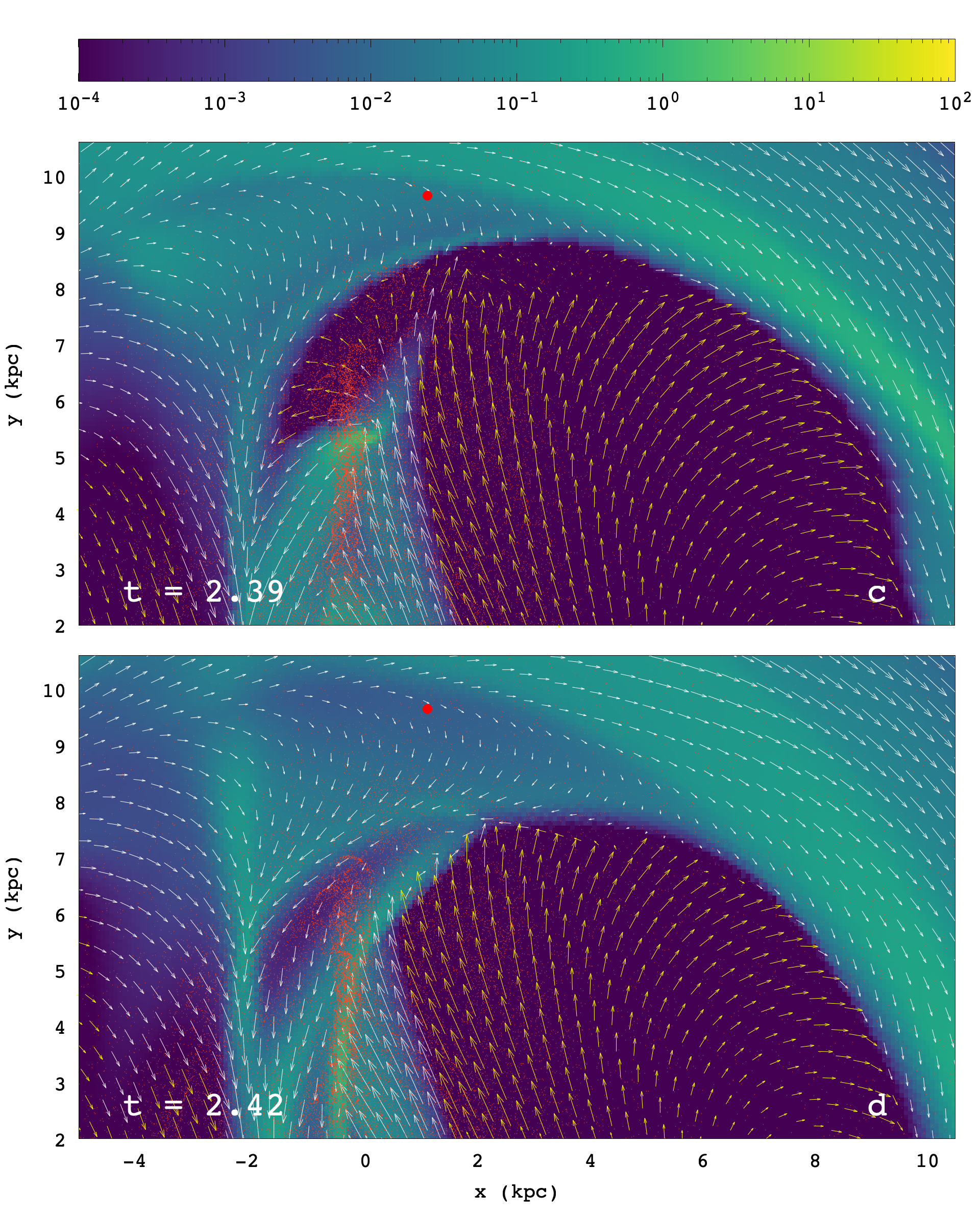}
\caption{(\textbf{a}--\textbf{d}): Characteristic evolution of the flow 
at the end and beyond the x1-bar region 
within a $\Delta t=0.11$ period, during which the extension towards the 
$\Gamma$-like feature breaks and forms again. Times are given at the lower 
left corner of the panels. The flow in the upper branch of the $\Gamma$
feature splits, pointing to two opposite directions at the points indicated 
with heavy yellow arrows in (\textbf{a},\textbf{b}). The long white arrow in (\textbf{a})
indicates the point ``A'', where the tail of the $\reflectbox{L}$ feature join 
the dust-lane shock in the bar region. Red dots mark the location of L$_1$.
\label{ram_stargas}}
\end{figure}
\begin{figure}[H]
\includegraphics[width=13.5cm]{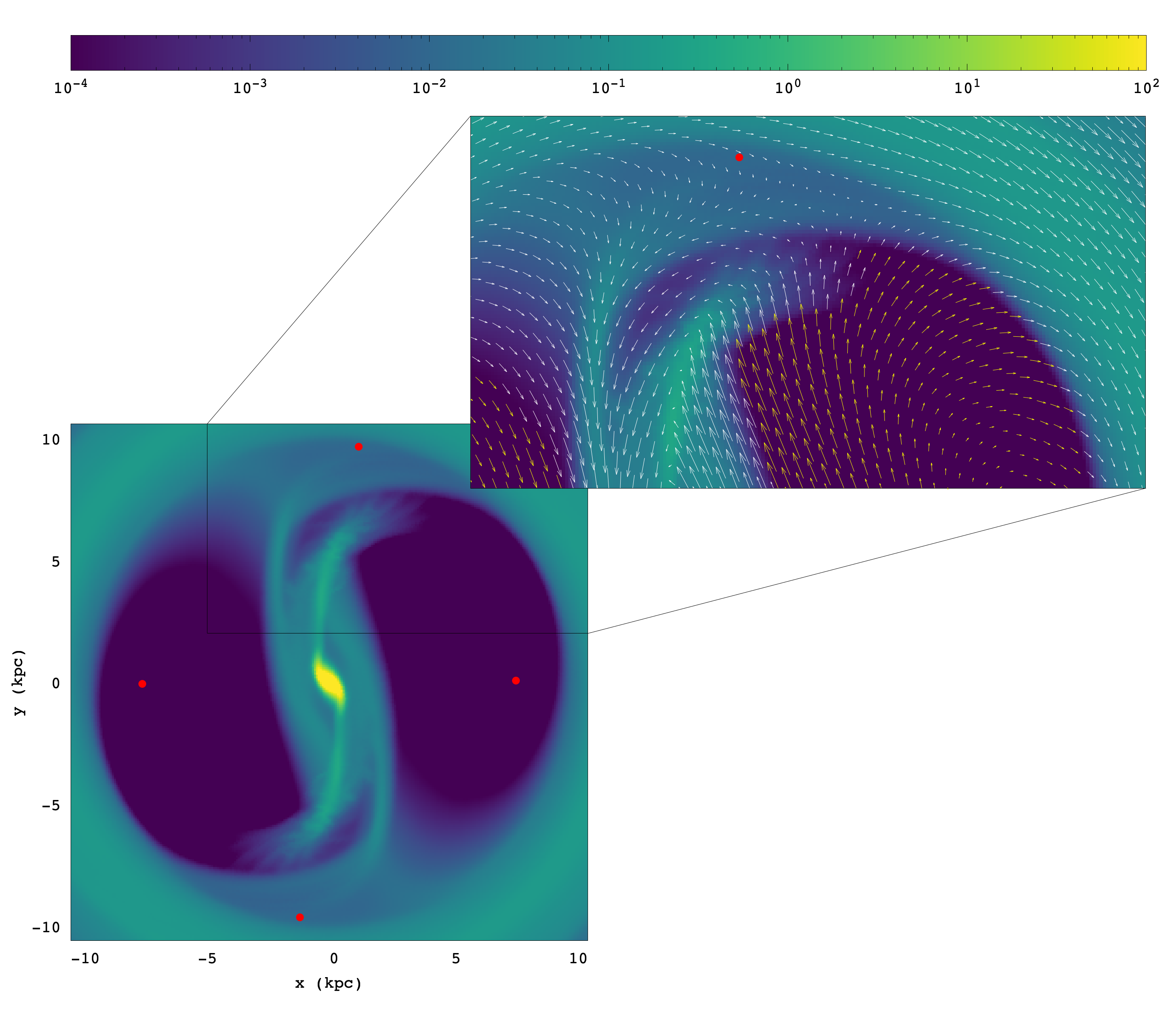}
\caption{A mean RAMSES response morphology, created by stacking the images of 
100 snapshots within 4.35 rotational periods of the system. It has all the 
typical morphological features discussed in the text. Red dots give the 
locations of the Lagrangian points.
\label{mean}}
\end{figure}
\begin{figure}[H]
\includegraphics[width=13.5cm]{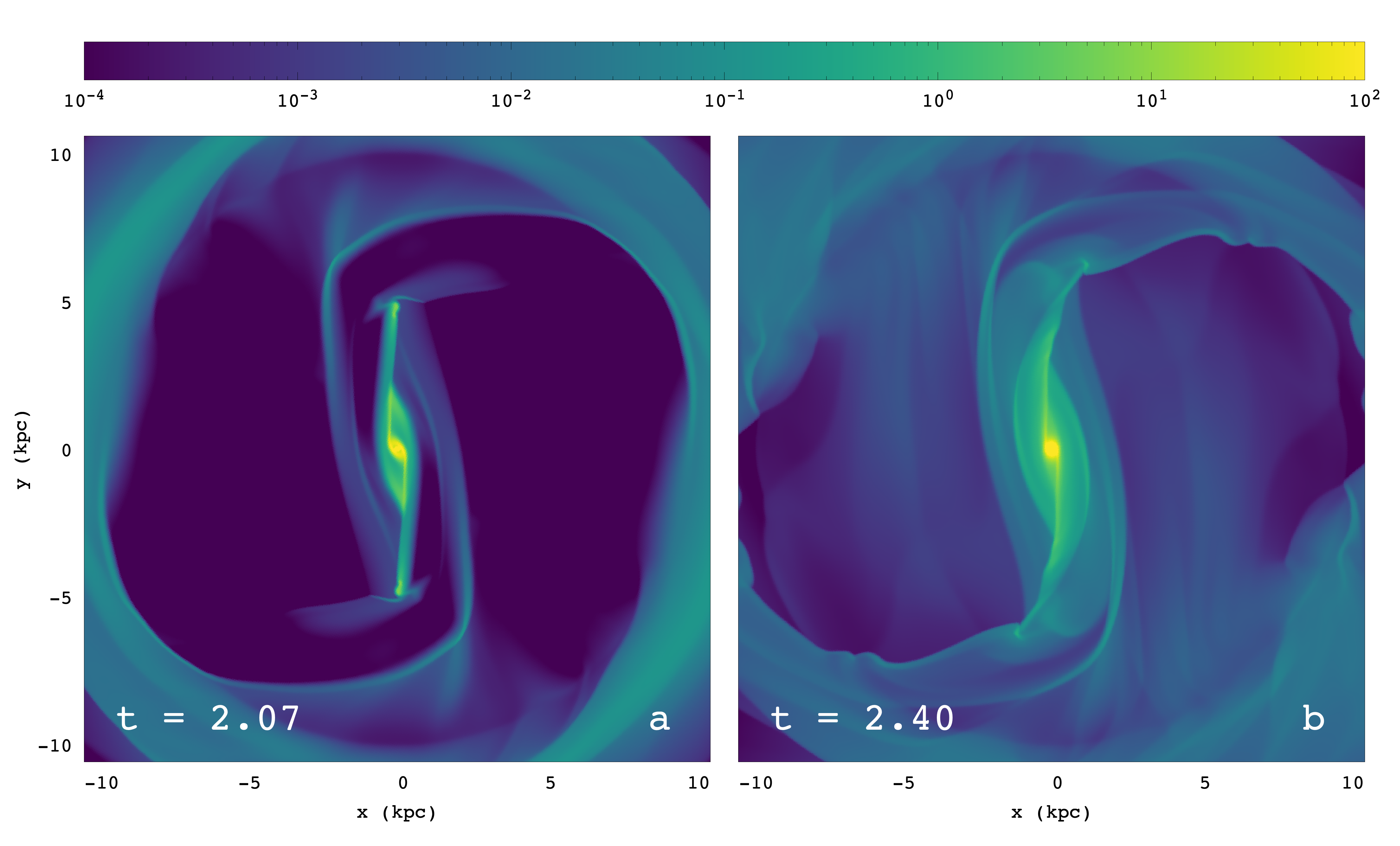}
\caption{Typical responses in a RAMSES model with $c_s=10$~km~s$^{-1}$ (\textbf{a}) and 
$c_s=20$~km~s$^{-1}$ (\textbf{b}). The rest of the initial conditions are as in the model 
of Figure~\ref{typicalss}. In (\textbf{a}), the response does not differ essentially from 
the model with $c_s=5$~km~s$^{-1}$, while in (\textbf{b}) we observe instabilities in the 
shocks in regions with high density gradients.
\label{ramsescs}}
\end{figure}

\section{Discussion}
\label{discuss}

\subsection{Code Dependence of the Responses}
\label{codcomp}

In our study, the equations of hydrodynamics were first derived in Lagrangian 
(co-moving) form, where the coordinate system is attached to moving mass 
elements, and then in Eulerian form, where the coordinate system is fixed in 
space. The corresponding codes used were versions of the SPH and RAMSES schemes, 
respectively.

Responses with both codes gave results with some common, main morphological 
features. These are: 1. The dust-lane shocks in the bar region, 2. the tails 
that are attached to them at the end of the straight-line part of the dust-lane 
shocks (indicated with arrows in Figures~\ref{typicalss} and  
\ref{ram_stargas}a), 3. the $\Gamma$-like features formed by 
these tails together with their extensions backwards with respect to the 
rotation of the system, and 4. the inclined ``x2'' region in the central part of the bar.

Nevertheless, there are several differences as well. They can be summarized by 
comparing Figure~\ref{sphrepl30} with Figure~\ref{mean}. Besides the different 
modeling techniques, we remind that, in SPH, we are removing from the simulation 
the gas particles that are found to be practically in escape trajectories, as 
well as half of the particles that are trapped in over-dense regions. The 
removed SPH particles were replenished by an equal number of particles placed 
randomly on the disk, with velocities corresponding to the circular velocity in 
the axisymmetric part of the potential. This was essential, because it allowed 
us to continue with a meaningful simulation at larger times. 
Nothing similar was  needed for the RAMSES simulations. On one hand, the SPH responses with the 
replenishment of the removed particles lead to more stationary models; on the 
other hand, the RAMSES results are more objective.

The most characteristic difference of the RAMSES snapshots compared with SPH
is the continuous displacement of the morphological features on the disk, 
mainly outside the bar region, something that leads to a cycle of the flow 
rather than to a stationary morphology. Nevertheless, there is a mean 
RAMSES morphology with clearly shaped structures (Figure~\ref{mean}). This mean 
snapshot and the corresponding mean flow are qualitatively similar with what 
we find in the SPH responses. Finally, another conspicuous difference is the 
low-density regions around L$_4$ and L$_5$ appearing in the RAMSES snapshots, 
mainly for $t > 3T$. This is not observed in the SPH models, due to the 
continuous replenishment of particles we applied (see also Section~\ref{svsg}
below).

\subsection{Stellar vs. Gas Response}
\label{svsg}

In both modeling techniques, there is a direct relation between the underlying 
orbital structure and the gaseous responses, and this relation is the same for 
the response models  of both the SPH and RAMSES codes. Namely, the straight-line 
dust-lane shocks are formed in the region where the x1 POs are elliptical-like with 
cusps (or tiny loops). In the particular potential we studied, these are x1 POs 
with projections on the $y$ axis of the system, reaching $y\approx 4.5$\,kpc. 
Due to the presence of the sine terms in the potential (Equation~(\ref{eq:potrthfour})), the 
x1 orbits slightly precess, facilitating, due to their orientation, the 
straight-line, dust-lane-shock formation (Figure~\ref{pos}a).

At larger distances from the center of the bar, the POs of this family develop 
loops of considerable size (Figure~\ref{orbsph}). This is a direct indication 
that the bar of the model is strong~\cite{gco88,a92a}. As soon as the POs of x1 develop 
conspicuous loops at their apocenters, the local density maxima of the gas are 
formed away from the x1-orbital flow. As the gas bypasses the x1 loops, it 
creates extensions of the dust-lane shocks, in general at an angle with them, in 
the direction away from the major axis of the bar.

The patterns that we call ``tails'' or ``bifurcations of the dust-lane shocks'' 
(indicated with arrows, e.g., in Figure~\ref{sphrepl30}b) appear at the sides of the 
bar. They are formed at the region where the ``banana-like'' flow around L$_4$ 
and L$_5$ meets the orbits of the f family. The dimensions of the longest 
rectangular-like, stable f POs marginally match the dimensions of a structure 
that can be vaguely described as boxy, formed essentially by two symmetric, 
with respect to the center of the system, $\Gamma$-like features 
($\Gamma$/\reflectbox{L}) (see, e.g., Figure~\ref{sphmod} and~\ref{typicalss}). 

As we can see in Figure~\ref{stellarespo}, the stellar bar, including its 
rectangular-like ``chaotic'' envelope as well, reaches a maximum 
$y\approx$ 8.5~kpc. This length coincides more or less with the $y$ at which 
we observe the ``horizontal'', roughly parallel to the $x$ axis, branch of the 
$\Gamma$ feature (Figure~\ref{mean}). In smaller radii from the center, we notice 
that the straight-line, dust-lane shocks, reach distances $y\approx$ 4.5--5~kpc. 
In the zone $5 < y < 8.5$~kpc, we have the x1 POs with loops and the orbits of 
the f family. However, the orbits of the f family, which reach distances 
6--7~kpc from the center, are unstable and, at even larger energies, they obtain 
6:1 type morphologies with loops (Figure~\ref{pos}b). They are characterized by 
sticky regions around them and along their unstable asymptotic curves 
(Figure~\ref{escsos}). An issue that needs further investigation is whether the 
shape of the extensions of the dust-lane shocks in the $5 < y < 8.5$~kpc zone 
and the formation of the ``horizontal'' branch of the $\Gamma$'s are determined 
just by the presence of the loops of the x1 orbits or if the asymptotic curves 
of the manifolds of the underlying unstable POs also play  a role.

The resulting morphologies are established essentially at the end of the 
growing phase of the bar that lasts for three rotational periods of the system. 
Numerical problems, such as the overdense clumps in SPH, or the transient 
``turbulent'' phase in RAMSES, occur close to the end of this period. Although 
the models survive in one or the other to longer times, understanding the 
smooth responses during the growing phase of the models remains an open 
question. Is it due to the growing character of the perturbation, or because 
the final amplitude of the model is large for real galaxies? In the latter case, 
response models could be used for estimating M/L ratios in near-infrared images 
of real galaxies.

Another interesting feature of the models is the low density in the regions of 
L$_{4}$ and L$_{5}$ at the sides of the bar. This is in agreement with the 
rather empty regions of the stellar responses and the fact that, for the $E_J$'s 
at which the banana-like orbits around L$_{4}$ and L$_{5}$ exist, non-periodic 
orbits are not trapped around them. They visit the neighborhood of L$_{1}$,
L$_{2}$ and then cross corotation. Such a behavior in SPH models with unstable 
POs around L$_{4}$ and L$_{5}$ has already been  found in a series of models in 
\cite{pt17}. The overall dynamics of the present case is consistent with the 
flows of those models.

\begin{figure}[H]
\includegraphics[width=13.0cm]{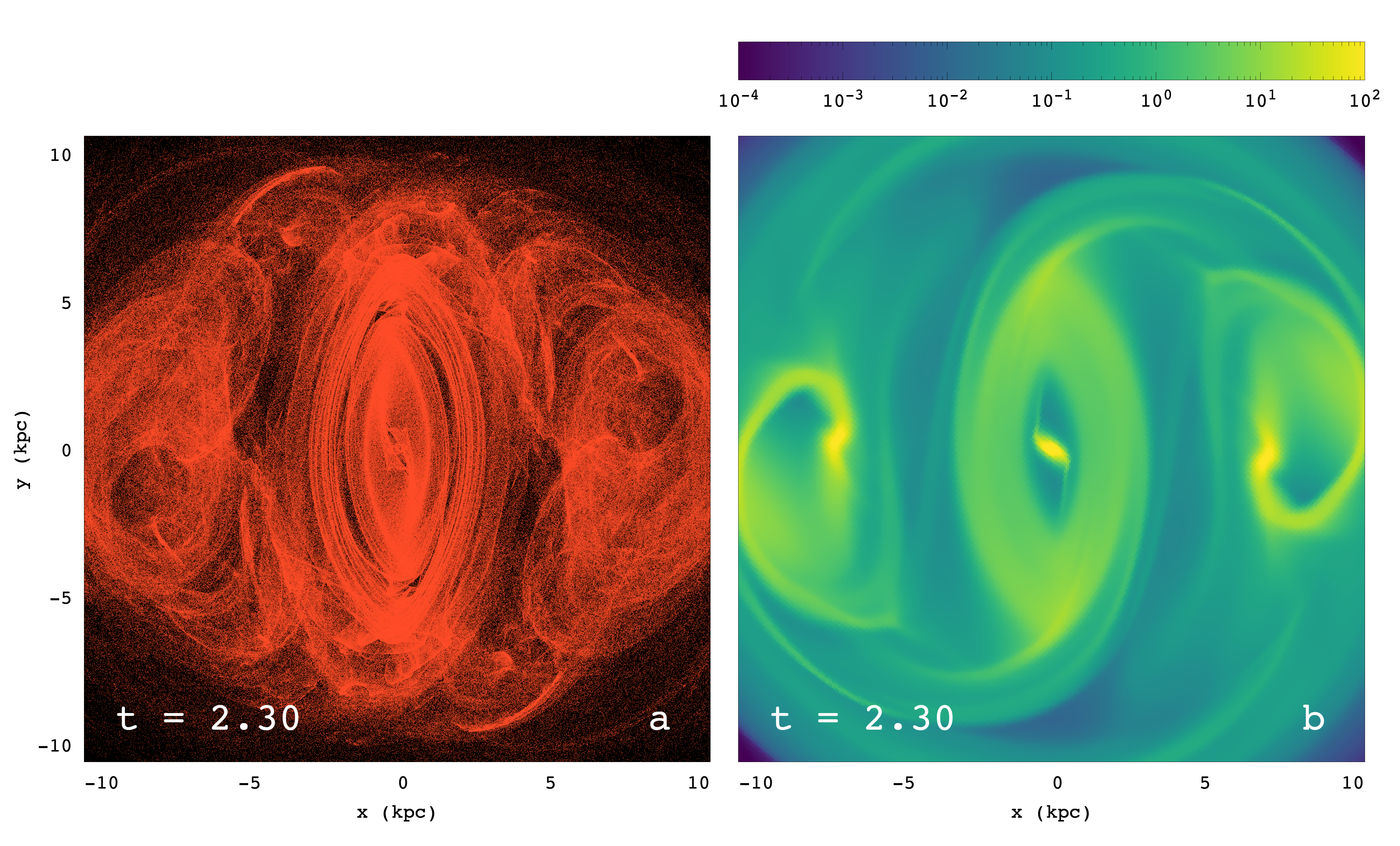}
\caption{A model in which the amplitude of the bar reaches only 50\% of that 
of the main model. (\textbf{a}) The stellar response. (\textbf{b}) The gaseous RAMSES model. The 
disks of the models are populated initially homogenously. The model has 
features pointing to a more regular orbital background.
\label{halfamp}}
\end{figure}

The instability of the banana-like and f orbits, in a considerable $E_J$ range, 
is typical for models with large bar perturbations. If we let the amplitude of 
the perturbation grow within $t=3T$ to 50\% of the maximum value used so far 
and then follow the response up to $t=10T$, stellar and gaseous responses 
change. The result is presented in Figure~\ref{halfamp} and the differences with 
the corresponding stellar (cf. Figure~\ref{halfamp}a with Figure~\ref{stellarespo}) 
and gaseous (cf. Figure~\ref{halfamp}b with Figure~\ref{mean}) responses are 
conspicuous. The disks with the initial conditions are 
homogenously populated, so it is evident that, in the case of 
Figure~\ref{halfamp}a, non-periodic orbits are now trapped around L$_{4}$ and 
L$_{5}$ and the gas density at the corresponding regions of Figure~\ref{halfamp}b 
are considerably larger than what we find in Figure~\ref{mean}. The L$_{4}$ and 
L$_{5}$ regions resemble those that are found in several models of specific 
barred galaxies \cite{treu12}. In addition, 
elliptical-like orbits reinforce a bar up to $y\approx$ 7~kpc in 
Figure~\ref{halfamp}a, while at the same region in Figure~\ref{halfamp}b we can 
observe a thick ring structure reaching this distance. The overall response in 
the bar region resembles the models in \citet{sbm15}.

In the present paper, we do not attempt to compare our models with features of 
specific galaxies, despite the fact that the model is the result of the 
estimation of a specific barred-spiral galaxy (NGC~7479). A systematic 
comparison of morphological features and of the form of dust lanes, with 
images of barred-spiral galaxies and the comparison of the associated velocity fields, 
will give valuable information about the underlying stellar orbital structure, the 
pattern speeds of the bars and the spirals, or even about the mass-to-light 
ratios on the disks, as has been performed in \cite{ff16}. This work is in progress.

\section{Conclusions}
\label{conclusions}

\begin{itemize}
\item The basic conclusion of the present study is that the dust-lane shocks 
in the gas responses of barred galaxies models avoid the regions in which the 
stable POs of the x1 family have developed sufficiently big loops. They deviate 
from these regions and, as they bypass them, they form extensions at an angle with the 
straight-line shocks. Ahead of them, in the direction of rotation, we find 
low-density regions, which in many cases have a ``triangular''-like shape, as, 
e.g., model 12 in \cite{a92b}. This morphology is encountered in both codes we used (SPH and RAMSES).

\item Responses during the growing-bar phase are smoother, in the sense that 
we observe fewer regions with strong density gradients. As the amplitude of the 
perturbation during this phase is always smaller than the final, maximum one, 
the models do not develop x1 POs with big loops (Figure~\ref{stellarespo}a). 
Thus, the straight-line, dust -lane shocks extend to larger distances.

\item A characteristic feature of the models are the ``tails'', which are 
dense regions that appear bifurcating from the straight dust-lane shocks at specific points 
along them. The bifurcating points are identified with the points at which the x1 POs 
start having considerable loops (indicated with arrows in Figure~\ref{sphrepl30}). 
Gas is streaming along these lanes towards the points where the three density enhancement join.

\item The SPH models can be followed for a long time only with replenishment 
of particles that are manually removed mainly from the overdense regions, but it
gives, after a certain time, an invariant response. RAMSES, on the other hand, 
has a short, relatively turbulent phase for $t \gtrapprox T_f = 3T$ and then reaches a 
repeating cycle, where the morphology of the straight-line dust-lane shocks and 
their extensions characterize the snapshots. Nevertheless, there is a 
dominating ``mean'' morphology, given in Figure~\ref{mean}.

\item Both codes give information valuable for understanding the dynamics of 
the model and should be used when comparison with the morphology of specific 
galaxies is attempted. With the Lagrangian SPH method, we can obtain detailed 
velocity fields in the the dense regions of the model, while with the grid code 
RAMSES, we can obtain an overall picture of the kinematics of the models. Both 
modeling techniques are useful for understanding gas dynamics in barred-spiral 
galaxies.

\item Finally, as regards the \textit{stellar} response, in this model, like in 
the case of the models for NGC~4314 we studied in \cite{paq97} and for 
NGC~1300 in \cite{pkg10}, we find a second bar, which is a ``chaotic'' envelope 
around the x1 bar. It is ``chaotic'', in the sense that it is supported by 
chaotic orbits. The consistency of its appearance in the models indicates that 
this is rather a common feature in barred galaxies.

\end{itemize}

\authorcontributions{Conceptualization, methodology, P.A.P. and E.A.; 
writing---original draft preparation, S.P. and P.A.P.; writing---review and 
editing, all authors; software and validation, S.P.; formal analysis, S.P.; data 
curation, S.P.; visualization, S.P.; supervision, P.A.P. and E.A.; project 
administration, P.A.P. All authors have read and agreed to the published version 
of the manuscript.}

\funding{This research received no external funding.}

\institutionalreview{Not applicable.}

\informedconsent{Not applicable.}


\dataavailability{The data underlying this article will be shared on reasonable 
request to the corresponding author.}

\conflictsofinterest{The authors declare no conflict of interest.} 



\abbreviations{Abbreviations}{
The following abbreviations are used in this manuscript:\\
\par\noindent
\begin{tabular}{@{}ll}
PO(s) & Periodic orbit(s)\\
SPH & Smoothed particle hydrodynamics\\
\end{tabular}}

%
%


\setenotez{backref=true,list-name={Note}}
\begin{adjustwidth}{-4.6cm}{0cm}

\printendnotes[custom]
 
\end{adjustwidth}

\begin{adjustwidth}{-\extralength}{0cm}

\reftitle{References}




\end{adjustwidth}
%


\end{document}